\documentclass[a4paper,10pt]{article}
\usepackage[pdftex]{graphicx,hyperref}
\usepackage{amsmath,amsfonts,amssymb}
\usepackage{fullpage}
\usepackage{authblk}
\usepackage{setspace}
\usepackage{subcaption}
\usepackage{booktabs}
\usepackage{hyperref}
\usepackage{tabularx}

\usepackage[explicit]{titlesec}
\usepackage{sidecap}
\usepackage{pbox}
\usepackage[superscript]{cite}

\usepackage{amsfonts}
\usepackage{amsmath}
\usepackage{multirow}
\usepackage{longtable}
\usepackage{graphicx}
\usepackage{xcolor}
\newcommand{\add}[1]{{\color{black}{#1}}}

\date{}
\title{\bf{From code to market: Network of developers and correlated returns of cryptocurrencies}}

\author[1,2]{Lorenzo Lucchini}
\author[3]{Laura Alessandretti}
\author[2]{Bruno Lepri}
\author[4]{Angela Gallo}
\author[5,6,7,*]{\\Andrea Baronchelli}
\affil[1]{University of Trento, Department of Information Engineering and Computer Science, Via Sommarive 9, Trento, 38123, ITALY}
\affil[2]{Fondazione Bruno Kessler, Via Sommarive 18, Trento, 38123, ITALY}
\affil[3]{Copenhagen Center for Social Data Science, University of Copenhagen, \O ster Farimagsgade 5, Copenhagen K, 1353, Denmark}
\affil[4]{Department of Finance, Cass Business School, London EC1Y 8TZ, UK }
\affil[5]{City, University of London, Department of Mathematics, London EC1V 0HB, UK}
\affil[6]{UCL Centre for Blockchain Technologies, University College London, UK}
\affil[7]{The Alan Turing Institute, British Library, 96 Euston Road, London NW12DB, UK}
\affil[*]{\small Corresponding author:  Andrea.Baronchelli.1@city.ac.uk}

\begin{document}
\maketitle
\vspace{1.5cm}
\textbf{``Code is law'' is the founding principle of cryptocurrencies. The security, transferability, availability and other properties of crypto-assets are determined by the code through which they are created. If code is open source, as customary for cryptocurrencies, this principle would prevent manipulations and grant transparency to users and traders. However, this approach considers cryptocurrencies as isolated entities, neglecting possible connections between them. Here, we show that \boldmath{4\%} of developers contribute to the code of more than one cryptocurrency and that the market reflects these cross-asset dependencies. In particular, we reveal that the first coding event linking two cryptocurrencies through a common developer leads to the synchronisation of their returns. Our results identify a clear link between the collaborative development of cryptocurrencies and their market behaviour. More broadly, they reveal a so-far overlooked systemic dimension for the transparency of code-based ecosystems that will be of interest for researchers, investors and regulators.}

\clearpage

	A cryptocurrency is a digital asset designed to work as a medium of exchange. The underlying Blockchain technology allows transactions to be validated in a decentralised way, without the need for any intermediary~\cite{nakamoto2008bitcoin}. Every cryptocurrency is entirely defined and governed by its code, which determines its security, functionality, availability, transferability and general malleability~\cite{antonopoulos2014mastering}. This ``code is law'' architecture immediately puts developers under the spotlight~\cite{lessig1999code}. Lack of transparency in the coding process might damage users and other stakeholders of the code~\cite{natl_law_rev}.  

``Open code'' is identified as the antidote to lack of transparency~\cite{lessig1999code}. Even if the code is accessible only to a small fraction of users, the reasoning goes, it would put the asset and stakeholders at repair from manipulations~\cite{lessig2007code}. For this reason, the code of the vast majority of cryptocurrencies is stored in public repositories. GitHub alone currently stores the code of more than $1,600$ cryptocurrencies~\cite{4_github_software_repository}.

Cryptocurrencies are nowadays used both as originally intended, i.e. media of exchange for daily payments and, to a larger extent, for speculation~\cite{ceruleo2014bitcoin,rogojanu2014issue}. 
The market value of a cryptocurrency is not based on any tangible asset, resulting in an extremely volatile, and largely unregulated, market~\cite{bitcoinrisks2020,gkillas2018application,chan2017statistical}. However, the cryptocurrency market  has attracted private and institutional investors~\cite{36_wide_attention_times,1_age_of_cryptocurrencies}. % 37_wide_attention_telegraph,38_wide_attention_cnbc,
At the moment of writing, more than $3,000$ cryptocurrencies are traded, capitalising together more than $200$ Billion dollars~\cite{cryptocompare2020page,CoinGecko2020}.

In this paper, we challenge the view that open code grants transparency to cryptocurrencies, even accepting that literate users do check it carefully (which is of course far from obvious). We do so by analysing $298$ cryptocurrencies (i) whose code is stored in GitHub and (ii) whose daily trading volume has been, on average, larger than $10^5$ USD~\cite{33_machine_learning_cryptocurrencies} during their lifetime. We show that:

\begin{enumerate}
	\item A substantial fraction of developers ($4\%$) contributes to the code of two or more cryptocurrencies. Hence, cryptocurrencies are not isolated entities but rather form a network of interconnected codes.
	\item The temporal evolution of the network of co-coded cryptocurrencies anticipates market behaviour. In particular, the first time two independent codes get connected via the activity of one shared developer marks - on average - a period of increased correlation between the returns of the corresponding cryptocurrencies.
\end{enumerate}

Thus, the temporal dynamics of co-coding of cryptocurrencies provides insights on market behaviours that could not be deduced based on the combined knowledge of the code of single currencies and the present state of the market itself. In other words, transparency, i.e., the availability of relevant market information to market participants, is a systemic property. The whole network of cryptocurrencies should be considered both by regulators and by professional investors aiming to maximise portfolio diversification. From this point of view, our work contributes a new dimension to the literature focused on the properties of the cryptocurrency market, which has so far adopted approaches ranging from financial~\cite{19_bubbles_in_bitcoin,18_liquidity_and_market_efficiency_in_cryptocurrencies,17_statistical_analysis_of_cryptocurrencies,20_market_for_cryptocurrencies,25_negative_bubbles_and_shocks_in_cryptocurrencies}, to behavioural~\cite{24_herding_in_cryptocurrencies}, from evolutionary~\cite{22_evolutionary_dynamics,23_predict_winner_in_market_cryptocurrencies,25_negative_bubbles_and_shocks_in_cryptocurrencies}, to technological~\cite{5_innovation_potential_in_cryptocurrencies,3_cryptocurrency_dataset} perspectives.

% network
\begin{figure*}[t]
	\centering
	\includegraphics[width=0.75\paperwidth]{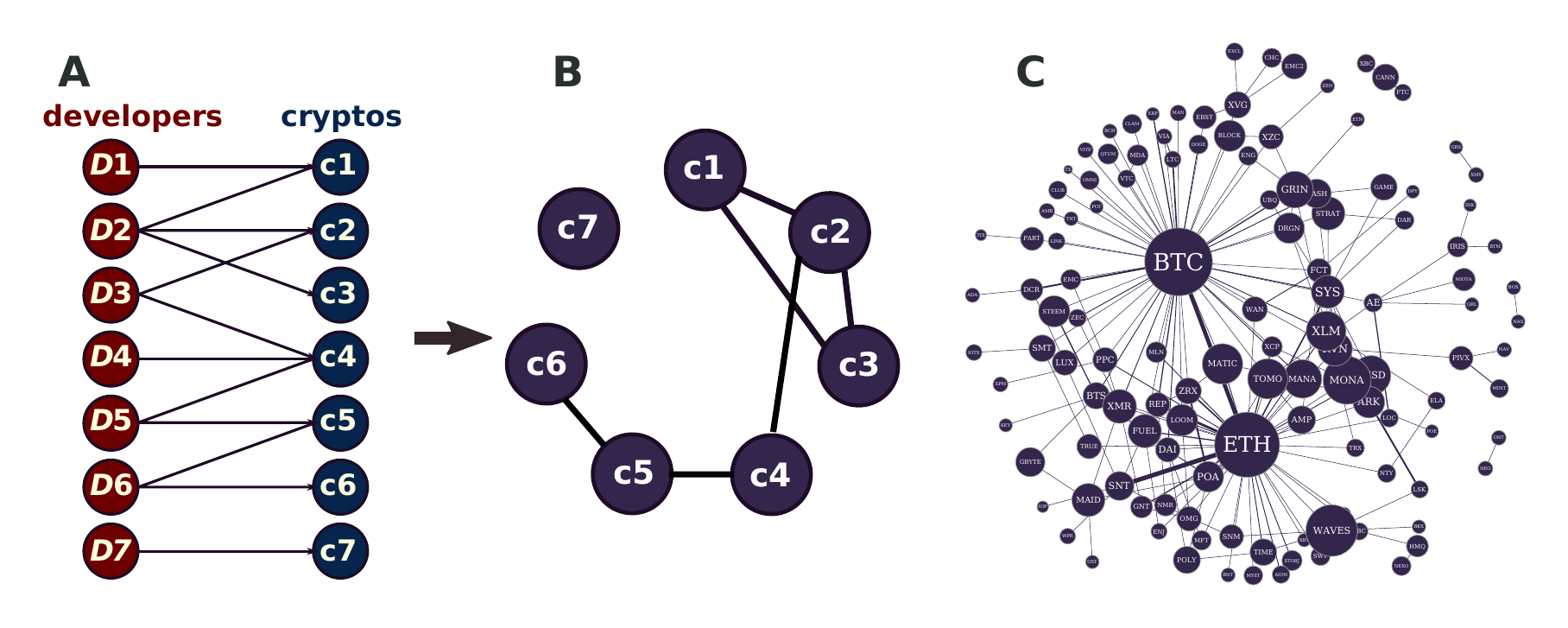}
	\caption{{\bf Fig. 1.}	\textit{The GitHub network of cryptocurrencies.} (\textbf{A}) The GitHub dataset can be represented as a bipartite network, where developers (red circles) are linked to the cryptocurrencies (blue circles) they have edited at least once. (\textbf{B}) Projection of the bipartite network, cryptocurrencies that have at least one common developer are connected. (\textbf{C}) The real network of $123$ cryptocurrencies with at least one connection. Node size is proportional to the number of connections and link width is proportional to the number of common developers between two cryptocurrencies. Bitcoin (BTC) and Ethereum (ETH) play a central role in the graph.}
	\label{fig1}
\end{figure*}

\section*{Results} \label{Results}

\subsubsection*{GitHub activity and the network of cryptocurrencies} We are interested in the coding and market activity concerning actively traded cryptocurrencies (see Materials and Methods). The $298$ cryptocurrencies with trading volume lager than $100,000$ USD whose code is stored on GitHub ($298$ projects) include $63$ out of the top $100$ cryptocurrencies, ranked by average market capitalization \add{during October 2019}. $6,341$ developers contributed to these GitHub projects, totalling $879,742$ edits (see SI \textit{Section} 1.2 for more details). The number of developers working on a cryptocurrency project correlates positively with its market capitalisation ($0.48$ Spearman correlation coefficient with p-value$<0.0001$, see Fig. S2 (A)), as previously noted~\cite{4_github_software_repository}. 

The activity of the developers is heterogeneous. $28\%$ of developers focused only on the top $10$ cryptocurrencies, producing $20\%$ of the edits, while only $15\%$ of the developers worked only on projects with a capitalisation lower than the median capitalisation of the market, producing only $11\%$ of the developing events. The \add{Ethereum} community soars above the others in terms of editing activity ($109,527$ development events), while Bitcoin has the largest number of developers, $832$ (Fig. S1). In general, the number of developers and the number of edits for a given project strongly correlate ($0.92$ Spearman correlation coefficient with p-value$<0.0001$, see Fig. S2 (B)).

We find that $4\%$ of developers contributed to more than one cryptocurrency, and are responsible for $10\%$ of all edits. We further investigate their role by representing the GitHub data as a bipartite network, where developers and cryptocurrencies (the nodes) are connected by edit events (the links) (Fig.~\ref{fig1} (A)). We then project the bipartite network, and obtain the network of connected cryptocurrencies where cryptocurrencies are nodes and a link exists between them if they share at least one developer (Fig.~\ref{fig1} (B)). We find that this network has $204$ links\add{, activated first by 147 different developers,} and $123$ non-isolated nodes, out of which $115$ form a giant component. Bitcoin has the largest number of connections, $53$, followed by Ethereum with $43$. The remaining $175$ projects do not share any developer (Fig.~\ref{fig1} (C)). The presence of \add{The presence of a small fraction of developers who contributed to more than two cryptocurrencies ($22$ out of $147$) makes the network rich in cliques} (see SI \textit{Section} 1.3 for more analyses on the network).

% work-scheme
\begin{figure*}[t]
	\centering
	\includegraphics[width=0.7\paperwidth]{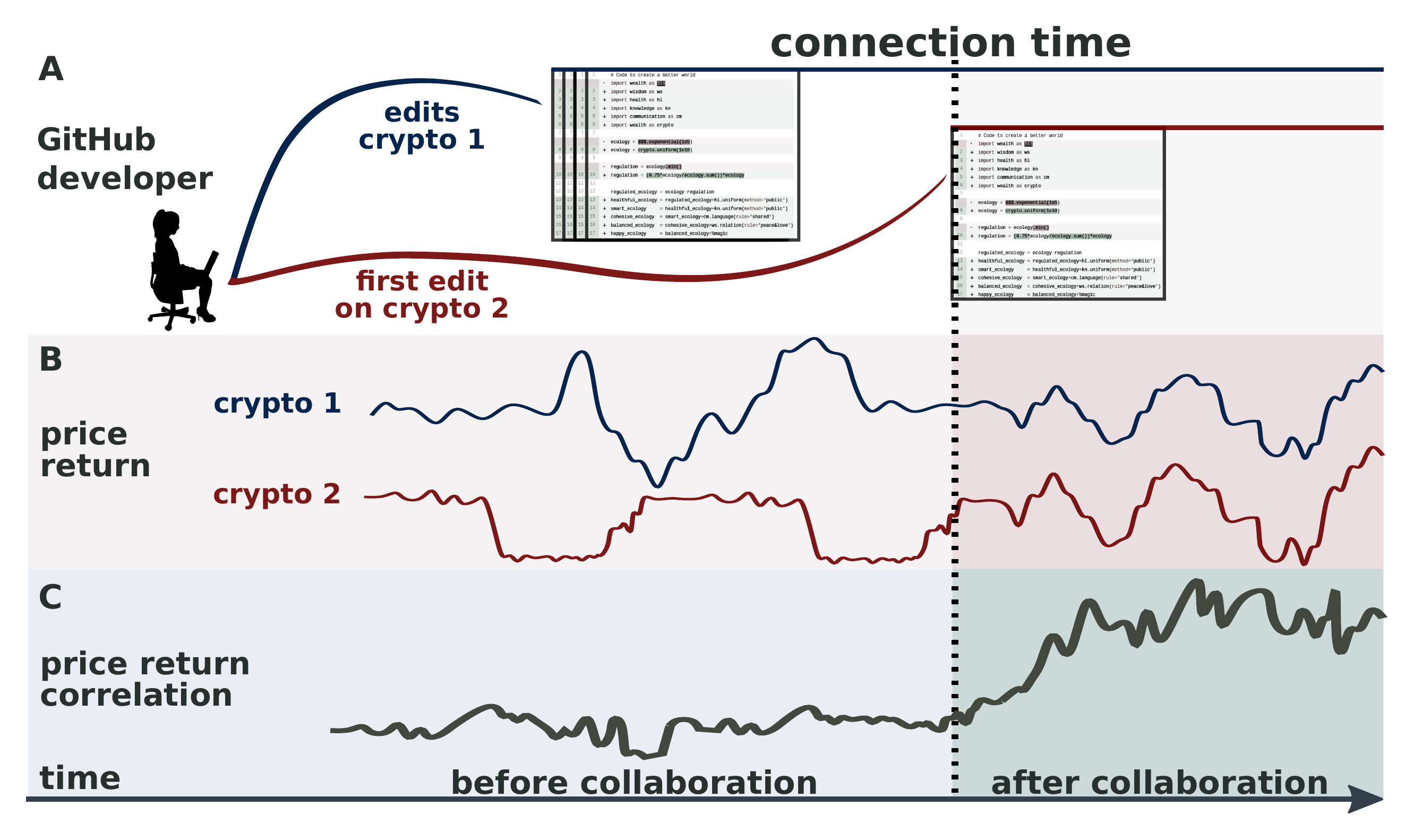}
	\caption{\noindent {\bf Fig. 2.} \textit{GitHub co-development and cryptocurrency market synchronisation.} (\textbf{A}) A developer of cryptocurrency ``crypto 1'' publishes her/his first contribution to ``crypto 2''. If no other developer has worked on both currencies before, this moment represents the GitHub \emph{connection time} for the pair composed of ``crypto 1'' and ``crypto 2'' . (\textbf{B}) The time-series describing the asset returns of the two currencies synchronise after the \emph{connection time}. (\textbf{C}) The Spearman correlation between the two time-series increases when the asset returns synchronise.}
	\label{fig2}
\end{figure*}

\subsubsection*{Market synchronisation of GitHub-linked cryptocurrencies} We now consider the temporal evolution of the cryptocurrency network over $5$ years of coding activity (from March $5$, $2014$ to May $30$, $2019$). A link between two cryptocurrencies is created the first time that a developer of one of the two edits the other (Fig.~\ref{fig2} (A)), referred in the following as the GitHub \emph{connection time}. What happens to the market behaviour of the two cryptocurrencies that have just been linked in the GitHub network?

\begin{figure*}[t]
	\centering
	\includegraphics[width=0.95\linewidth]{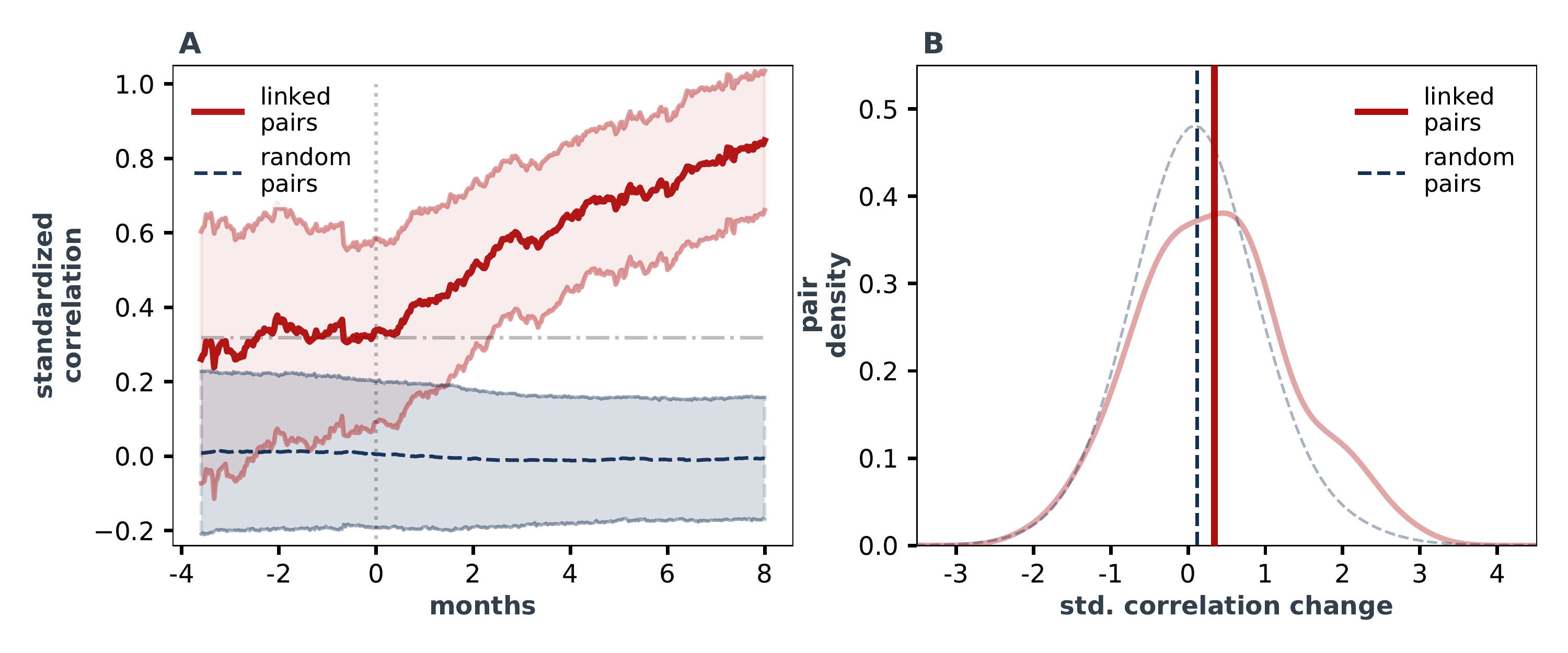}
	\caption{{\bf Fig. 3.} \textit{Market synchronisation following GitHub connection time.} (\textbf{A}) Average standardized Spearman coefficients between return time-series of linked pairs (red line) and \add{ a sample of random pairs of cryptocurrencies (dashed-blue line). The size of the random samples is chosen to be the same as the number of existing linked pairs at each time. Its average size in the period reported in Fig. \ref{fig3}(A) is $124$.} Shaded areas represent $2$ standard \add{deviation} of the mean and are determined via bootstrap (see Sec. Methods). The grey dot-dashed line corresponds to the average standardized correlation in the $3$ months before the connection occurred. Time is shifted such that $d=0$ corresponds to the GitHub \emph{connection time} of each pair. Correlations are measured over a $4$-month rolling window.
	(\textbf{B}) Distributions of the average correlation for linked and random pairs. Averages are computed over periods of four months: the four months before the \emph{connection time} and the period between $2.5$ and $6.5$ months after the \emph{connection time}. Vertical lines correspond to the average of each distribution. Pairs that synchronised after the \emph{connection time} shift the distribution towards positive values. All the density distributions are computed using a Gaussian Kernel Density Estimation \add{setting the bandwidth values to $0.39$.} For raw data histograms see Fig. S12.}
	\label{fig3}
\end{figure*}

We focus on the correlation between asset returns~\cite{8_stability_of_mst_of_pricereturn_volatility, 9_networks_equities_in_financial_markets}. We rescale time so that the \emph{connection time} corresponds to $d=0$ for each pair of GitHub-linked currencies and we measure the Spearman correlation over a \add{backward} rolling window of size $s=4$ months (see Fig.~\ref{fig2} (B and C), Fig.~\ref{fig3} (A), and Materials and Methods for definitions; results are robust with respect to variations of this definition, see SI \textit{Section} 1.4.1). To limit the effect of overall changes in market evolution, we standardize the value of the Spearman correlation, for a given pair of linked currencies and at a given time, by subtracting the average correlation across all possible pairs of currencies at that time and dividing by the corresponding standard deviation (see Materials and Methods).

Fig.~\ref{fig3} (A) shows that the average standardized Spearman correlation between the returns of two linked cryptocurrencies\add{, averaged over the set of 204 linked pairs,} increases at the turn of the GitHub \emph{connection time}, rising from $0.31 \pm 0.01$, on average \add{($\pm$ standard error of the mean)}, in the four months before the \emph{connection time}, to $0.66 \pm 0.01$, in the period included between $2.5$ and $6.5$ months after the \emph{connection time} (Fig.~\ref{fig3} (A), significant under Welch-test\cite{44_avg_test_welch,45_welch_test_ruxton}, with p-value $p=0.02$). This corresponds to a relative increase of almost $130\%$ after the synchronization occurred (see SI \textit{Section} 1.9.2 for details about the synchronisation period). This results is robust to major perturbations of the network, including the removal of Bitcoin or\add{/and} Ethereum from it (Fig. S9).

We test that the observed behaviour is specific to linked pairs by measuring the synchronisation of a random sample of $10^4$ cryptocurrency pairs, selected from the entire market excluding linked pairs. Their \emph{connection time} is chosen at random from the list of actual GitHub \emph{connection time}s (SI \textit{Section} 1.4.1 for different randomisation approaches). 
We find that the standardised correlation of such pairs remains constant across the \emph{connection time}, ruling out the possibility of ecology effects induced by the specific distribution of \emph{connection time}s (Fig. S5). 
We note also that, on average, the standardised Spearman correlation is higher for linked pairs compared to random pairs.

The increase in correlation observed for linked pairs could (i) be driven by few outliers or (ii) reflect the behaviour of the majority of them. Fig.~\ref{fig3} (B) shows the distributions of the increase in standardised correlation between the $4$ months preceding and the $4$ months included between $2.5$ and $6.5$ months after the \emph{connection time}. The distribution of linked pairs is centred at positive values of change (i.e., increase in correlation) and shows a significantly higher average synchronization compared to the distribution of random pairs, e.g., under Welch test (for more statistical tests see SI \textit{Section} 1.4.1). In particular, approximately $65\%$ of linked couples increased their correlation after GitHub \emph{connection time}, a percentage significantly higher than random (Fig. S13). These observations confirm that the observed change in correlation is not simply driven by outliers, hence supporting hypothesis (ii).

The market behaviour of cryptocurrencies is also characterised by other properties. We repeated the analyses reported above to study the correlations between the time-series describing daily changes in trading volume and market capitalisation. We found no significant effects of the \emph{connection time} on those measures (see results in SI \textit{Section} 1.8) under a Welch test at significance level $0.05$.

% transition &  Distributions before after

\subsubsection*{Market properties of GitHub-linked cryptocurrencies}
\begin{figure}[!t]
% Volume and market differences
	\centering
	\includegraphics[width=0.65\linewidth]{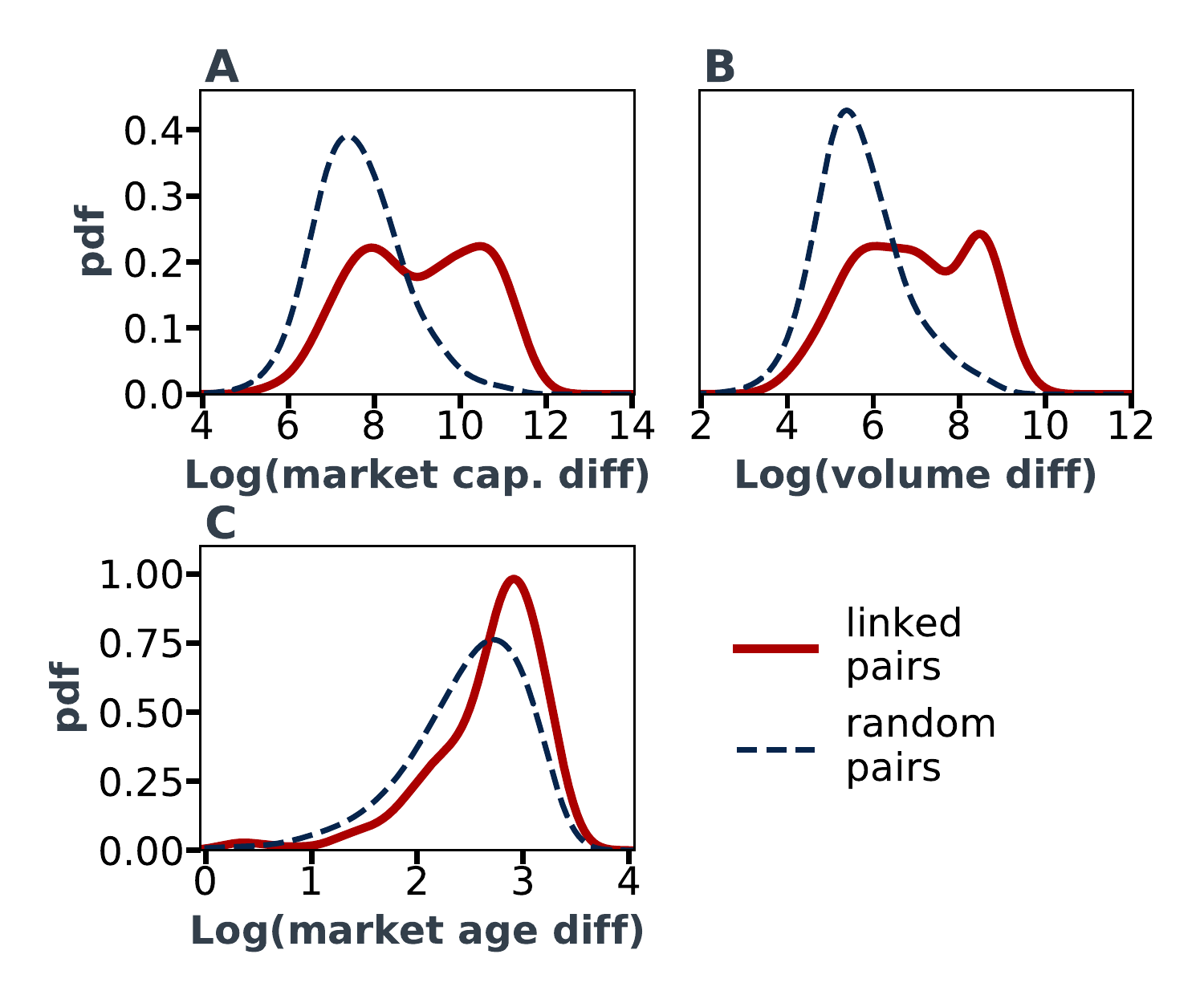}
	\caption{{\bf Fig. 4.} \textit{Linked pair composition.} (\textbf{A}) Probability density function of the difference in market capitalisation among cryptocurrencies forming linked pairs (continuous line) and random pairs (dashed line). (\textbf{B}) Probability density function of the difference in transaction volume among cryptocurrencies forming linked pairs (continue line) and random pairs (dashed line). (\textbf{C}) Probability density function of the difference in market age at the connection time among cryptocurrencies forming linked pairs (continue line) and random pairs (dashed line). \add{All the density distributions are computed using a Gaussian Kernel Density Estimation setting the bandwidth values to $0.36$.}}
	\label{fig4}
\end{figure}

We now consider the market properties of GitHub-linked cryptocurrencies across GitHub \emph{connection time}. First, we focus on the difference in market capitalization and volume among pair constituents. We find that the absolute difference in market capitalisation and volume between two linked cryptocurrencies is typically larger than between randomly selected cryptocurrencies (see Fig.~\ref{fig4} (A), Fig.~\ref{fig4} (B), and SI \textit{Section} 1.9.2 for details; note also that the market capitalization and volume of currencies are highly correlated, as expected (Fig. S17).

Then, we shift our attention to differences in market age, defined as the difference in the amount of time since a currency appeared in the market. We find that the age difference of the two cryptocurrencies in a linked pair, measured at \emph{connection time}, is significantly higher, on average, than the difference of market age observed for random pairs (Fig.~\ref{fig4} (C)). In particular, we find that the second-edited currency is younger than the first-edited currency in $61\%$ of the cases, and has lower market capitalization in $65\%$ of the cases.

Finally, we investigate the factors responsible for the observed heterogeneity in synchronization across linked pairs (Fig.~\ref{fig3} (B)). We find that, when a linked pair includes one of the top-$10$ linked cryptocurrencies in terms of market capitalization (evaluated in the period preceding \emph{connection time}), the corresponding synchronization of returns following connection is significantly higher than average (Fig. S21 (D)). Other factors, including the type of development event (push or pull), the direction of the link (from younger to older or vice-versa), and the \emph{connection time}, do not explain the observed differences in synchronization across pairs (Fig. S21 (A-B, E-F)). 

\section*{Discussion}

We analysed the relationship between code and market for $298$ GitHub-hosted cryptocurrencies whose trading volume was larger than $10^5$ USD for the covered period. We showed that approximately $4\%$ of developers contributed to the code of more than one cryptocurrency and that these developers are more active than the average, contributing together to $10\%$ of all edits. We then defined the network of co-developed cryptocurrencies and showed that, for months after the GitHub \emph{connection time}, the correlation between the return time series of two GitHub-linked cryptocurrencies increased, on average. We found that other market indicators - and in particular volume - do not show the same behaviour. Finally, we showed that developers tend to work on an established currency first and that linked pairs containing at least one top cryptocurrency exhibited a larger correlation of returns following connection.

It is important to delimit the scope of our findings. First, we only considered projects developed on GitHub. While this is by far the largest repository of \add{cryptocurrency open-source code (it hosts more than $99\%$ of the project hosted on online repositories)}, alternatives exist, e.g., \emph{GitLab}~\cite{gitlab2020}. Second, we selected cryptocurrencies based on their average trading volume, possibly neglecting currencies with only a short history of significant trading volume. 
Third, we focused on the first connecting event and did not investigate the presence and consequence of a possibly increasing pool of shared developers between two cryptocurrencies and/or actions of the developer(s) in that pool.  Fourth, we considered pairs of cryptocurrencies, neglecting other possible influences of the network built in the first part of the paper. Finally, we did not consider the structure of the code or the semantics of the coding that a developer of the first cryptocurrency performs on the second. All these are open directions for future work.

\add{Of course, our analysis can not identify the mechanisms that drive the observed market synchronisation. Speculatively, at least two main dynamics might be at play. The first identifies code as an important "fundamental" for this market \cite{kristoufek2015main,li2017technology}. Traders would be aware of and operate (also) based on code and code development. The activity of developers would therefore represent a signal that, perceived by many traders, could result in the observed synchronization.  The second dynamics - either complementary or alternative to the previous one - points to a greater role for developers, who could either directly own and trade large amounts of the cryptocurrencies they edit or be hired by stakeholders who in their turn do the trade. At the systemic level, these interlocking directorates of developers/stakeholders would cast a shadow on the transparency of the market and potentially expose it to systemic risks due to hidden structural correlations between cryptocurrency prices. 
	%Here, signals in the coding space would be produced and perceived by a small fraction of privileged traders.
	In this respect, it is worth noting that the lack of incentives for developers is a long-standing issue for cryptocurrencies. Some Bitcoin developers, for example, are paid by companies with an interest in Bitcoin~\cite{bitcoin_coders}, in the case of Ethereum some are funded by the Ethereum Foundation itself while bug-bounties, development grants and visibility remain other common incentives \cite{voshmgir2019token}. In this context, our results could suggest that trading on the cryptocurrency market might play the role of incentive for developers to perform certain cross-currency actions. Interestingly, the lack of increase in synchronisation for volumes suggests that the observed synchronisation of returns is not due to an overall increase in trading interest towards the linked cryptocurrencies. Beyond these two mechanisms, more explanations may exist and exhausting or testing them - if at all possible - is outside of the scope of this paper.}

Our results have broad implications.  Code has become an important societal regulator that challenges traditional institutions, from national laws to financial markets~\cite{lessig2007code,de2018blockchain,16_cryptocurrencies_and_blockchain}. In particular, whether and how financial markets and technological - code - development interact is an open and debated question~\cite{4_github_software_repository,5_innovation_potential_in_cryptocurrencies,6_bitcoin_effects_in_return,7_cryptocurrencies_as_financial_asset}. The case of cryptocurrencies is paradigmatic and still largely unexplored. Cryptocurrencies are open-source digital objects traded as financial assets that allow, at least theoretically, everyone to directly shape both an asset structure and its market behaviour. Our study, identifying a simple event in the development space that anticipates a corresponding behaviour in the market, establishes a first direct link between the realms of coding and trading. In this perspective, we anticipate that our results will be of interest to researchers investigating how code and algorithms may affect the non-digital realm~\cite{mayer2009can,yuan2019data,ali2019discrimination} and spark further research in this direction.

\section*{Materials and Methods}

\subsection*{Data}
\subsubsection*{The GitHub dataset}\label{subsec:Methods_github_data}

GitHub is a service providing a host for software development using Git version control system~\cite{39_github_introduction_bell,40_github_perils_kalliamvakou} largely used in a variety of innovation fields, from science to technological development~\cite{26_science_on_github}. Previous research on the platform focused on the understanding of collaborative structures and developer behaviour, showing the importance of social characteristics in the selection of code modifications~\cite{27_contribution_in_github} and of socialisation as a precursor of joining a project~\cite{28_social_technical_factors_in_github}.

A project is stored on GitHub in a so-called ``repository'', and its production-ready code lives in the ``master branch'' of the repository~\cite{githubGuides2020}. Developers can modify the master branch in two ways, depending on their role. So-called ``collaborators'' are part of the core development team, and can directly edit the code by triggering a ``push event''. In contrast, ``contributors'' are anyone who contributed some changes to a project, by submitting their suggestions through a ``pull request'' that was later accepted and merged by one of the ``collaborators''. Thus, ``push'' and accepted ``pull requests'' are the core events in the development of cryptocurrency production-ready code ~\cite{29_social_coding_in_github}.

We retrieved cryptocurrency GitHub repository names from CoinMarketCap~\cite{coinmarketcap}. We find that $1668$ out of the $2225$ cryptocurrencies listed in CoinMarketCap as of $9$ June $2019$ shared their source code on GitHub. Then, we queried the GitHub Archive dataset~\cite{githubarchive}, that stores all events on public repositories from $2011$, through GoogleBigQuery~\cite{GoogleBigQuery}. This step provided us with all events related to the development of cryptocurrency GitHub projects. Specifically, we queried two types of events: ``push events'' and accepted ``pull request
events''. Finally, we removed all events triggered by GitHub apps (software designed to maintain and update the repositories), and we removed from our dataset GitHub profiles whose name included the term ``bot'' not to include noise from users that identified or were reported to be non-human. 

\subsubsection*{The market dataset}

We collected cryptocurrency daily price, exchange volume and market capitalisation from three different web sources: CoinGecko~\cite{CoinGecko2020}, CryptoCompare~\cite{cryptocompare2020page} and CoinMarketCap~\cite{coinmarketcap} (the latter only until the end of July $2018$ due to updates in the website regulations). We processed and compared the data following the suggestions by Alexander and Dakos~\cite{41_cmc_cg_cc_cryptodata_alexander}. Discrepancies of CryptoCompare daily values from the CoinGeko ones, if larger than $500\%$, were discarded and treated as missing values.

The price of a cryptocurrency represents its exchange rate (with USD or Bitcoin, typically) which is determined by the market supply and demand dynamics. The exchange volume is the total trading volume across exchange markets, in dollars. The market capitalisation is calculated as a product of a cryptocurrency circulating supply (the number of coins available to users) and its price. We retrieved historical data for currently inactive currencies by querying all the $6,000$ and more cryptocurrencies recorded in the CoinGeko database~\cite{CoinGeckoAll}. Our datasets include market indicators from April $3$, $2013$ (date by which all the webpages started collecting data), until October $30$, $2019$. Note that, to study the effects of Github development on market indicators, we collected market data for $6$ months longer compared to the Github data.

In this work, we focus on cryptocurrencies that can be traded with sufficient ease. We therefore consider only cryptocurrencies whose trading volume is larger than $100,000$ USD~\cite{33_machine_learning_cryptocurrencies}. We find that $520$ cryptocurrencies meet this condition (see Table S1 for full list), out of which $297$ share their code on Github. 

%We refer to those cryptocurrencies that survived the filtering process as widely traded.

\subsection*{Randomized pairs} \label{subsec:randomize}
We compare various quantities measured for GitHub-linked pairs to the corresponding values measured for random pairs. A random pair is obtained by (1) extracting two of the $520$ cryptocurrencies that meet the condition of an average daily market volume larger than $100,000$ USD and (2) verifying that the two extracted cryptocurrencies do not form together a GitHub-linked pair. As for the average volume, days with zero transaction volume (days of market inactivity) were discarded and treated as missing values.

\subsection*{Randomized pairs} \label{subsec:randomize}
We compare various quantities measured for GitHub-linked pairs to the corresponding values measured for random pairs. A random pair is obtained by (1) extracting two of the $521$ cryptocurrencies that meet the condition of an average daily market volume larger than $100,000$ USD and (2) verifying that the two extracted cryptocurrencies do not form together a GitHub-linked pair. As for the average volume, days with zero transaction volume (days of market inactivity) were discarded and treated as missing values. \add{The resulting set of $521$ cryptocurrencies represent $27\%$ of all the cryptocurrencies with a market history on both CryptoCompare and CoinGecko.}
%This approach is intended to include, in the set of cryptocurrencies under study, also cryptocurrencies that are presently inactive and no longer traded in the market. }

\subsection*{Time-series analysis} \label{subsec:Methods_crypto_activity}

A cryptocurrency asset return at time $t$ is defined as $\textstyle{R(t)=\frac{P(t)-P(t-1)}{P(t-1)}}$, where $P(t)$ is the price~\cite{30_volatility_in_finance}. The change in market capitalisation at $t$ is defined as $\textstyle{C_{M}(t)=\frac{M(t)-M(t-1)}{M(t-1)}}$, where $M(t)$ is the market capitalisation. The change in volume is defined as $\textstyle{C_{V}(t)=\frac{V(t)-V(t-1)}{V(t-1)}}$ where $V(t)$ is the volume as time $t$.

Following a standard approach in time series analysis~\cite{micciche2003degree,press1992numerical}, we measure correlation as the Spearman coefficient between two time series. To compare the correlation across pairs of currencies, following, e.g., Schruben~\cite{43_timeseries_standardization_schruben}, we compute the standardized correlation as

$$SC_{k}(t) =  \frac{{C_{k}}(t)- \bar{C}(t)}{\sigma (t)},$$

\noindent where $C_{k}(t)$ is the the correlation time series, computed for a pair $k$ by comparing the return time series ($R_{i}(t)$ and $R_{j}(t)$) of paired assets $i$ and $j$ at time $t$, and $\bar{C}(t)$ and $\sigma (t)$ are the average correlation and corresponding standard deviation across pairs. \add{At each time step t, the set of pairs used to compute the standardized correlation consists of the pairs for which we had price data at time t.}

%\add{We approached the analysis of these time-series by means of different consolidated non-parametric tests. Specifically, we tested the differences between the averages of standardised correlation distributions by means of a unequal variance t-test (Welch-test)~\cite{44_avg_test_welch, 45_welch_test_ruxton}. 

\subsection*{Error estimate and bootstrapping}\label{subsec:Methods_bootstrap} We compute the error associated with the average standardised correlation across pairs using bootstrapping~\cite{chernick2011bootstrap}. \add{For each value of $d$, representing the number of days before/after the connection time (such that at the connection time $d=0$): (i) we sample Nd pairs of currencies with replacement, where $N_d$ is the number of existing linked pairs at time $d$;} (ii) we compute the average standardised correlation $SC(d)=\sum_{k=1}^{N_d}SC_{k}(d)/N_d$ where $k$ is running across the $N_{d}$ pairs; (iii) we repeat steps (i) and (ii) $10^4$ times; and (iv) we compute the mean and standard deviation across the obtained values of $SC(d)$. These values provide an estimate of the average standardised correlation and associated error for the population of linked pairs \add{$d$ days after the connection}. We follow the same procedure for random linked pairs.

\bibliographystyle{Science}

\section*{Acknowledgments}
The authors are grateful to Abeer ElBahrawy for support with data collection and many useful conversations.

\section*{Data availability}
All data needed to evaluate the conclusions in the paper are present in the paper and/or the Supplementary Information (see SI \textit{Section} 2).

\section*{Author contributions}
A.B. conceived of the project; L.L., L.A., B.L., A.G. and A.B. designed research; L.L. and L.A performed research; L.L., L.A., B.L., A.G. and A.B. analysed the data and discussed results; L.L., L.A., B.L., A.G. and A.B. wrote the paper.

\section*{Competing interests statement}
The authors declare no conflict of interest.

%Here you should list the contents of your Supplementary Materials -- below is an example. 
%You should include a list of Supplementary figures, Tables, and any references that appear only in the SM. 
%Note that the reference numbering continues from the main text to the SM.
% In the example below, Refs. 4-10 were cited only in the SM.     
\cleardoublepage

\section*{Supplementary materials}
Supplementary Text\\
Data and code availability\\
Table S1\\
Figs. S1 to S21\\
\vspace{1.5cm}

% For your review copy (i.e., the file you initially send in for
% evaluation), you can use the {figure} environment and the
% \includegraphics command to stream your figures into the text, placing
% all figures at the end.  For the final, revised manuscript for
% acceptance and production, however, PostScript or other graphics
% should not be streamed into your compliled file.  Instead, set
% captions as simple paragraphs (with a \noindent tag), setting them
% off from the rest of the text with a \clearpage as shown  below, and
% submit figures as separate files according to the Art Department's
% instructions.

\bibliographystyle{unsrt}
\bibliography{preprint_bib}

\subsection{List of cryptocurrencies and their market rank}
In Table 1 we report the list of cryptocurrencies studied in this work. For each cryptocurrency, we analysed the corresponding market behaviour. The left-hand side of each column reports the rank of the $521$ cryptocurrencies based on their market capitalization as of October $30$, $2019$. The right-hand side of each column reports the symbol corresponding to each cryptocurrency project. We report in \textbf{bold} the projects hosted on GitHub, and in \textbf{{\color{red} bold-red}} linked cryptocurrency projects.

\subsection{GitHub activity} \label{sec:SI_crypto_developers}

In this section, we describe the activity of cryptocurrency developers. The data were collected as follows. First, we retrieved the set of $521$ cryptocurrencies with average daily transaction volume greater than $10^5$ USD from the CryptoCompare and CoinGeko websites between April $3$, $2013$, and October $30$, $2019$. We found that, by the beginning of June $2019$, $298$ projects out of $521$ are developed on GitHub. Then, we retrieved the source code page for each of them in CoinMarketCap and retrieved their GitHub activity via the GoogleBigQuery API. We considered only GitHub events occurred before June  $1$, $2019$, to guarantee that our market data cover at least $150$ days of market activity, after that date, for all active assets.

Public cryptocurrency projects on GitHub are edited by organizations of GitHub developers. Developers can contribute in different ways, but only a few types of contributions shape the structure of a project, i.e. ``PushEvent'' and accepted (i.e. ``merged'') ``PullRequestEvent''. In Fig. S1, we show the number of different event types and the number of users triggering the events. Panel (A) reports the top $20$ cryptocurrencies in terms of the number of events triggered (left panel) and the number of developers (right panel). The \emph{Ethereum} community is the largest. Panel (B) (top) shows the number of events triggered on GitHub cryptocurrency projects sorted by event type. Panel (B) (bottom) shows the number of developers triggering an event sorted by event type. We find that ``PushEvents'' and ``PullRequestEvents'' are among the most common events (alongside with ``IssueCommentEvents''), but are executed by a relatively small amount of users, e.g. compared to ``WatchEvents''.

% Activity and developers per Cryptocurrency project

We find a negative correlation (-0.80 Spearman coefficient with p-value$<0.001$) between the number of developers who triggered a certain number of events, and the number of such events, implying that most developers are not very active while few developers are responsible for a larger number of events. This relation is well captured by a power-law fit \textit{number\_of\_events} $\sim $\textit{number\_of\_developers}$^\alpha$ with exponent $\alpha = -1.11$. We also find a positive correlation between the number of developers working on a cryptocurrency and the number of developing events triggered by the project ($0.92$ Spearman coefficient with p-value $<0.001$).

Finally, we focus on the $\sim4\%$ of all developers who worked on more than one project. We find a significant difference between the average number of events $N_{e}$ (or median $M_{e}$) for the developers working on one-crypto,  $N_{e}=174$ ($M_{e}=30$), compared to the developers working on two cryptos, $N_{e}=375$ ($M_{e}=100$), by means of a Welch test (t-statistic $= -3.01$, with p-value $= 0.003$).

We confirm previous results on the positive correlation between the amount of activity of project developers and the respective cryptocurrency market capitalization (\textit{51}). In particular, we find that the Spearman correlation between the two is $0.49$, with p-value $< 0.001$, (Fig. S2 (A)). The positive correlation between capitalization and number of developers is even stronger ($0.72$, with p-value $< 0.001$) when the median market capitalization per group of cryptocurrency with the same number of developers is considered. Figure S2 (B) shows the strong correlation between the number of developers and the number of development events ($0.92$ Spearman correlation, with p-value $< 0.001$).

% Activity and developers per Cryptocurrency project

\subsection{The structure of the network}\label{subsub:SI_network}
In this section, we describe the properties of  \textit{linked cryptocurrencies}, defined as pairs sharing at least one developer, who contributed at least one ``PushEvent'' or one accepted ``PullRequestEvent'' to both projects. The data was collected by selecting accepted ``PullRequestEvents'' by parsing the ``payload'' field of the GitHub Archive dataset and looking for pull requests that were later merged into the project. 

The relations among linked pairs are encoded in a network using the procedure presented in Fig. 1 of the main paper.     

\add{The network is constructed by the work of $147$ unique developers, which are responsible for the first activation of $204$ links between $123$ different projects. Among these developers, a few ($22$) are responsible for more than one connection. In Fig. S3 we report the number of developers as a function of the connections they are responsible for. The inset graphs show the highly connected communities of GitHub projects created by the top three developers in terms of activated connections.  We note that no project is shared between the two developers responsible for five connections, named ``GitterBadger'' and ``JeremyRubin''. The same two users share only one project each with developer ``EvertonMelo'' (who is responsible for $19$ connections). Importantly, the common projects are Ethereum and Stellar, two among the top-10 projects based on the number of developers active on GitHub.}

\add{Here we focused on the network as a whole (see Fig. 1 (C)). Below we present some of its properties.}
We find that the node degree distribution, measured at the beginning of June $2019$, is well described by a power-law function, with exponent $-1.3$ (see S4 (A)). Also, we find that the number of new linked pairs over time is well characterized by exponential growth (see Fig. S4 (B)). Fits are performed using a standard Markov Chain Monte Carlo (MCMC) process, where the error is estimated from a normally distributed prior. Finally, we find that the network is non-assortative, since its assortativity coefficient, $a = 0.07\pm 0.02$ is approximately zero.

% Activity and developers per Cryptocurrency project

In Fig. S5, we classify the cryptocurrency pairs based on the event that triggered the connection (``PushEvent'' or ``PullRequestEvent''). It is interesting to note that, during the last year, the number of collaborations originated by a ``PullRequestEvent'' has grown, surpassing the number of connections originated by a ``PushEvent'', i.e. by direct modifications of the code.

% New connection per event type

\subsection{Robustness tests}

In this section, we provide detailed information about how we computed correlations and the definitions we adopted. We show that the results presented in the main text are robust to significant variations of these definitions.

\subsubsection{Robustness with respect to window size} 
\label{subsec_SI:sliding_robust_and_random}
There are two widely used measures of correlations: the Pearson correlation and Spearman correlation coefficients. The first coefficient assumes that the data is normally distributed around its average and is used to test whether there is a linear relation among the data. In contrast, the Spearman correlation coefficient does not require normality. Thus, when dealing with financial time-series, it is common practice to use the second (\textit{41}). Computing the Spearman correlation coefficient for time-series data involves the choice of a time window over which comparing the two series. The correlation values typically depend on the size of this window. Smaller windows capture short-term variations, while longer ones capture long-term variations. In Fig. S6, we show the synchronization of pairs at the turn of the connection time, for different window sizes included between $60$ days and $180$ days. We find that the standardised correlation of linked pairs (red continuous line) grows after connection time under all these choices of the window.

% Robustness window size (S6)

\paragraph{Hypothesis tests.}
\label{subsec_SI:robustness_before_after}
The increase in correlation for linked pairs significantly differs from randomly sampled pairs for all window sizes under several tests: the Welch test, the Mann-Whitney U test, the Kruskal-Wallis and Kolmogorov-Smirnov tests (at significance level $\alpha=0.05$). 
As an exception, the Kolmogorov-Smirnov test fails to reject the null hypothesis (with $\alpha$ threshold $0.05$) for windows of size $160$ (with p-value $0.065$) and size $200$ (with p-value $0.069$). This results are reported in Fig. S7.

% Distributions of changes under differnt window sizes (S7)

\subsubsection{Robustness with respect to the randomization} \label{subsec_SI:robustness_randomization}
We introduce different types of randomizations. 
We define as \textit{RT random pairs} (random pairs with random connection time), pairs such that each of the components is randomly sampled among the entire ecology. This type of randomization is the one adopted as a comparison to the linked cryptocurrencies in the main text. We define as \textit{RTA random pairs} (random pairs sampled from cryptocurrencies with similar market age with a random connection time) pairs selected as follows: For each actual connected pair, we select two currencies from the ecology with a similar age at connection time (compared to currencies forming the original pair, $\pm7 days$). We define as \textit{ORTA random pairs} (random pairs sampled with one cryptocurrency taken from connected cryptocurrencies and the second sampled from cryptocurrencies with similar market age with a random connection time), pairs selected as follows: For each actual connected pair, we keep one of the two currencies and we select a second currency with similar age compared to the original second currency in the pair.
We find (see Fig S8) that ORTA pairs are significantly more correlated than RT pairs, while RTA random pairs do not significantly differ from RT random pairs. The results presented in the main text are computed using random pairs of type RT. 

% Multiple randomizations (S8)

\subsubsection{Robustness with respect to linked pairs subsampling} \label{subsec_SI:robustness_subsampling}
The two most central nodes in the collaboration network of cryptocurrency development are Bitcoin (BTC) and Ethereum (ETH).
each of them collects about $25\%$ of the total links of the network. To prove that the synchronization of linked pairs is not merely driven by connection with these two cryptocurrencies we subsample the set of linked pairs excluding, first, linked pairs including BTC (see Fig. S9 (A)), and, second, linked pairs including ETH (see Fig. S9 (B)).
A mild decrease in the amplitude of the correlation is observed in both cases, whilst significant difference is maintained (shaded areas in Fig. S9 represent $2$ standard deviations of the average).

\add{Interestingly, when simultaneously excluding from the network those pairs composed by either Bitcoin or Ethereum the average correlation before the connection goes to zero while the increase after the connections maintain a similar amplitude (see Fig S9 (C)).}

% Robustness under subsampling (S9)

\add{Note that our results are robust also with respect to the strength of the connection between two currencies. In particular, we found that there is a significant increase in standardised correlation following the first connection event, even when we exclude from the analysis pairs of currencies that connect more than once, i.e. there are multiple co-developers shared by the pair (see Fig S10).}

% Robustness under subsampling (S10) one_link!

\subsection{Median correlation}
Figure 3 in the main manuscript displays the average standardised correlation of linked pairs and random pairs (continuous red line and dashed blue line respectively). Instead, in Fig. S11, we compute the median of each set of pairs. The continuous red line reports the median correlation of linked pairs bootstrapped following the procedure presented in Materials and Methods (main text). Similarly, the dashed blue line shows the median value of the standardised correlation for the random pairs. Shaded areas represent the [$0.025$, $.0975$] quantile regions region estimated from the median distribution obtained from the bootstrapping procedure (\textit{60}). Dot-dashed grey line represent the median value of the linked cryptocurrencies before the connection time. One can notice that the synchronisation starts about three months later than in Fig. 3 but, in both cases, the maximum correlation is reached $9.75$ months after the connection time. The transition towards synchronization is more sharp when results are computed for the median. 

% Median Standardised correlation (S11)

\subsection{Correlation change raw histograms}
Figure S12 reports the raw histograms for the standard correlation change of asset returns. Continuous red line shows the raw distribution for linked pairs. Dashed-blue line shows the raw distribution for random pairs (RT pairs). 

% histograms of correlation change (S12)

\subsection{Increasing correlation}
In this section, we analyse the fraction of linked pairs whose standardised correlation increases after the connection time.
Figure S13 shows the fraction of linked pairs (continuous red line) with an increase in their standardised correlation against random pairs (dashed blue line). On the x-axis, we report the size of the window on which the standardised correlation is averaged.
The average fraction (on the y-axis) is computed following a bootstrap procedure as presented in Materials and Methods. Shaded areas represent $2$ standard deviations of the averages and are determined via bootstrap procedure (see Sec. Methods). Results show a significantly higher fraction of pairs whose standardised correlation increases after the connection time.

% increasing fraction of pairs (S13)

\subsection{Correlation of other measures}\label{SI:corr_other_measure}

In the main text, we discussed how the returns of linked pairs of currencies synchronize. Here, we describe the synchronisation of other market indicators centred at the connection time. Figure S14 shows the standardised correlation for the following measures and changes: (A) transaction volume, (B) daily change of transaction volume, (C) market capitalisation 
(D) daily change in market capitalisation, (E) price, (F) volatility. In particular, we calculated the price volatility as $v(t)=\ln(Ph(t))-\ln(Pl(t))$, where $Ph(t)$ and $Pl(t)$ are the highest and lowest daily value of the price, respectively (\textit{61},\textit{62}). The correlation was then computed over the $v(t)$ time series as for other measures.

% correlation of other metrics (S14)

The market indicators in the left column of Fig. S14 are non-stationary, thus the correlations of linked pairs do not significantly differ from the random pairs. \add{We tested that the averages of the two correlation distributions of  market capitalizations and volumes does not change significantly, following a connection event, using the Welch test.}
Instead, the indicators presented in the right column are stationary, allowing to discern interesting behaviours. We find that the daily change in volume (Fig. S14, (B)) and volatility (Fig. S14, (F)) are significantly more correlated for linked pairs even before the connection time. In both cases, there is a mild positive trend to synchronisation up to $50$ days before the connection time. In contrast, linked pairs do not significantly differ from randomized pairs when comparing their daily change in market \add{capitalization}(Fig. S14, (D)).

\subsection{Pairs and market}\label{subsub:SI_couplesDistributions}

In this section, we analyse the properties of linked pairs, by comparing them with a random sample of $95,451$ cryptocurrency pairs. This sample is drawn by selecting: (i) two random cryptocurrencies from the whole set of $521$ currencies and (ii) a connection time from the distribution of connection times (see SI \textit{Section} 1.4.2 for more details).

\subsubsection{Age at connection} \label{paragraph_SI:age_connection}
While GitHub activity of a linked cryptocurrency project started before or at its connection time, the same does not hold for their market history. In fact, the development history produced by connecting developers proceeds separately from the market series. Some cryptocurrencies might already be established currencies in the market while some of their developers only joined recently the developing team. Conversely, developers might start working on GitHub projects before a cryptocurrency is priced and traded in the market. At the time of their connection we can, thus, study this mismatch by looking at the market age of the oldest (the currency that has been for a longer time in the market among the two in a pair) and youngest cryptocurrencies in a pair. As reported in Fig. 4 (C), we find that there is a larger difference in age between the oldest and youngest cryptocurrency composing a pair compared to what observed in random pairs. Here, in Fig. S15, we show that both old and young linked cryptocurrencies systematically show an average higher age at the connection time than random pairs (Welch test and KS test at significance value $0.001$). However, it is important to note that, while the age distribution of young linked cryptocurrencies peaks around $0$, old linked cryptocurrencies have a significantly higher age, peaking around $800$ days. Being the cryptocurrency ages taken at the connection time, this result better explain a feature of the connection dynamic: old cryptocurrencies and young ones connects one another around the time at which the youngest ones are entering the market.

% Age distribution collaborative and random (S15)

We further investigate the role played by market age for linked cryptocurrencies by comparing the age (Fig. S16 (A)) and age-difference (Fig. S16 (B)) distributions for cryptocurrencies and pairs of currencies whose correlation increases and decreases after connection time. To this extent, only pairs with non-null correlation before and after the connection are taken into account.
We find a non-significant difference between these two classes (through both Welch test and KS test at a significance level of $\alpha=0.05$). This result suggests that effects related to market age do not affect the synchronization of returns for linked cryptocurrencies.

% Age difference (S16)

%%%% Market and volume differences
\subsubsection{Market and volume differences} \label{paragraph_SI:market_volumdiff}

In this section, we study the market capitalization and transaction volume of linked cryptocurrencies. The market capitalization is a proxy for cryptocurrency value, and transaction volume is a proxy of its usage. The two measures are found to strongly correlate (see  Fig. S17). In particular, the correlation is higher for oldest (and thus more established) cryptocurrencies, with a $0.91$ Spearman correlation coefficient (with p-value $\ll 0.001$), than for youngest ones, with a $0.60$ Spearman correlation coefficient (with p-value $\ll 0.001$).
Moreover, in Fig. S17, we also find that youngest linked cryptocurrencies tend to have lower transaction volume and market capitalisation compared to oldest ones (through Welch and K-S tests at a significance level of $0.001$).

% Market and volume (S17)

We then analyse the market capitalization and volume of linked pairs. In Fig. S18, we show the distribution of the absolute difference between two currencies in a pair, in terms of market capitalisation (panel (A)) and transaction volume (panel (B)). We find that, for linked pairs (dashed and continuous lines), the distributions span a broader range of values compared to random pairs (dot-dashed and dotted lines), both in terms of capitalisation and transaction volume. In contrast, we find no significant difference between linked pairs before (dashed lines) vs after (continuous lines) the connection time (under a KS two-sample test at significance level $\alpha=0.05$). Here, by value ``before" the connection we mean the average value in the $120$ days preceding the connection, while for ``after" we mean the average in the period included between $75$ days and $195$ after the connection. This choice accounts for the existence of a transition period following the connection (see Fig.~3 in the main text), whose length is found by measuring the point of maximum growth of the synchronization, by fitting the synchronization curve with a sigmoid function via a standard MCMC technique.

% Mrcp and trvol differences distribution (S18)

We further investigate market differences between linked pairs and random pairs, by comparing pairs whose correlation increases or decreases after the connection. Increasing pairs are defined as pairs whose average standardised correlation in the period after the connection, is larger than the average correlation computed before the connection. We find no significant difference between the two concerning their market indicators (under KS test at a significance level of $0.05$, see Fig.~ S19). We verified that the difference between the behaviour observed for pairs with increasing and non-increasing synchronization is non-significant independently of when it is measured, before or after the connection time (see Fig. S20).

% Mrcp differences distribution with increasing couples (S19)

% Trvol differences distribution with increasing couples (S20)

\subsubsection{Which pairs of linked cryptocurrencies synchronize?}

In the main text, we observed that when cryptocurrencies connect on GitHub, their returns tend to synchronize. This behaviour, however, is observed only for some observed connections. In this section, we investigate how various characteristics of the currencies involved in the connection affect whether or not they display synchronization. 

Our procedure works as follows. Based on given characteristics of the pair (see below) we divide linked pairs into two classes of pairs. Then, we test (under the following tests: Welch's test for average, Mood's and Kruskal-Wallis for the median, Mann-Whitney U test, and Kolmogorov-Smirnov test, at significance level $\alpha=0.05$) if there is a significant difference between the two classes, with one of the two showing a higher increase in correlation following the connection time. Results are shown in Fig. S21.\\

We focus on the following characteristics: 
\begin{itemize}
	\item {\textbf{Interplay between market age and order of development. } The first class (y-f) includes pairs such that the oldest currency in the pair is also the first-edited. The second class (y-s) vice-versa, see Panel (A).}
	\item {\textbf{Difference in market capitalisation between pair components}. The two classes include pairs whose components have a high difference in market capitalisation (high-diff) vs pairs whose components have a lower difference (low-diff) than the median market capitalisation difference, see Panel (B).}
	\item {\textbf{Developer behaviour following the connection}. The first class includes cases when a developer stops working on a currency in favour of another (dismiss), while the second includes cases where a developer continues working on both currencies (keep), see Panel (C).}
	\item {\textbf{Market capitalization of the linked currencies}. The first class (top10) includes pairs with at least one of the top $10$ currencies (in terms of market capitalization), while the second (minor) includes currencies that are not among the top $10$, see Panel (D).}
	\item {\textbf{Type of link}. The first class (Push) includes pairs connected via a ``PushEvent''; the second class includes pairs connected via an accepted ``PullRequestEvent'' (Pull), see Panel (E).}
	\item {\textbf{Time of connection}. The first class (late) includes the  $50\%$ more recent connections, and the second class (early) the other ones, see Panel (F).}

\end{itemize}

The hypothesis that the two classes display the same increase in correlation can be rejected in two cases only. First, under the \textit{keep-dismiss} classification (Panel C), we find that the correlation increases more when the common developer dismisses the first project compared to when he/she keeps working on it. Second, under the \textit{top10-minor}, we find that when one of the top-$10$ cryptocurrencies is included in the linked pair, there is a higher chance that the pair returns will synchronize (Panel D). 

% Synch in different ecologies (S21)

\section{Data and code availability}
The raw data analysed during this study are publicly available, upon free registration to the services, from \href{https://bigquery.cloud.google.com/}{GoogleBigQuery} platform and from \href{https://coinmarketcap.com/api/}{CoinMarketCap}, \href{https://min-api.cryptocompare.com/}{CryptoCompare}, and \href{https://www.coingecko.com/en/api}{CoinGecko} APIs.
Processed data allowing to reproduce the findings of this study are available in figshare (with the identifier 10.6084/m9.figshare.12994190). 
The code used for the analysis is made available at \url{https://github.com/LLucchini/FromCodeToMarket/}.

%%%%%%%%%%%%%%%%%%%%%%%%%%%%%%%%%%%
%%%%%%%%%%% TABLES %%%%%%%%%%%%%%%%
%%%%%%%%%%%%%%%%%%%%%%%%%%%%%%%%%%%
\clearpage
\section{Tables}

\noindent {\bf Table S1.} \textit{List of cryptocurrencies under study and their market rank}. The left-hand side of each column reports the rank of the $\mathbf{521}$ cryptocurrencies under study based on their market capitalization at October $\mathbf{30}$, $\mathbf{2019}$. The right-hand side of each column reports the symbol corresponding to each project. In \textbf{bold} we report the projects hosted on GitHub and in \textbf{{\color{red} bold-red}} linked cryptocurrency projects.
\begin{table}[h!]
	\label{tab:crypto_list}
	\centering
	\resizebox{1.025\textwidth}{!}{
		\begin{tabular}{rl}
			\toprule
			rank &                          sym \\
			\midrule
			1 &    \textbf{{\color{red}BTC}} \\
			2 &    \textbf{{\color{red}ETH}} \\
			3 &                 \textbf{XRP} \\
			4 &    \textbf{{\color{red}BCH}} \\
			5 &                         USDT \\
			6 &    \textbf{{\color{red}LTC}} \\
			7 &                          EOS \\
			8 &                          BNB \\
			9 &                 \textbf{BSV} \\
			10 &    \textbf{{\color{red}ADA}} \\
			11 &    \textbf{{\color{red}XLM}} \\
			12 &    \textbf{{\color{red}TRX}} \\
			13 &    \textbf{{\color{red}XMR}} \\
			14 &   \textbf{{\color{red}LINK}} \\
			15 &                           HT \\
			16 &  \textbf{{\color{red}MIOTA}} \\
			17 &                          OKB \\
			18 &                          XTZ \\
			19 &                         ATOM \\
			20 &   \textbf{{\color{red}DASH}} \\
			21 &    \textbf{{\color{red}NEO}} \\
			22 &                          ETC \\
			23 &    \textbf{{\color{red}ONT}} \\
			24 &                          CRO \\
			25 &                          MKR \\
			26 &                 \textbf{XEM} \\
			27 &   \textbf{{\color{red}DOGE}} \\
			28 &                          BAT \\
			29 &    \textbf{{\color{red}ZEC}} \\
			30 &                 \textbf{PAX} \\
			31 &                          VET \\
			32 &   \textbf{{\color{red}TUSD}} \\
			33 &   \textbf{{\color{red}QTUM}} \\
			34 &    \textbf{{\color{red}ZRX}} \\
			35 &    \textbf{{\color{red}DCR}} \\
			36 &                          HOT \\
			37 &    \textbf{{\color{red}RVN}} \\
			38 &                 \textbf{BTG} \\
			39 &                \textbf{VSYS} \\
			40 &    \textbf{{\color{red}OMG}} \\
			41 &                         LUNA \\
			42 &    \textbf{{\color{red}LSK}} \\
			43 &                \textbf{NANO} \\
			44 &                         ALGO \\
			45 &    \textbf{{\color{red}BTM}} \\
			46 &                          KCS \\
			47 &                 \textbf{DGB} \\
			48 &                          BTT \\
			49 &    \textbf{{\color{red}REP}} \\
			50 &                 \textbf{BCD} \\
			51 &               \textbf{THETA} \\
			52 &    \textbf{{\color{red}DAI}} \\
			53 &  \textbf{{\color{red}WAVES}} \\
			54 &                           SC \\
			55 &                 \textbf{ICX} \\
			56 &   \textbf{{\color{red}MONA}} \\
			57 &    \textbf{{\color{red}BTS}} \\
			58 &                 \textbf{BCN} \\
			59 &                 \textbf{KMD} \\
			60 &   \textbf{{\color{red}MAID}} \\
			61 &     \textbf{{\color{red}AE}} \\
			62 &                          MXM \\
			63 &                 \textbf{QNT} \\
			64 &                         IOST \\
			65 &    \textbf{{\color{red}XVG}} \\
			\bottomrule
		\end{tabular}
		\begin{tabular}{lr}
			\toprule
			rank &      crypto\\
			\midrule
			66 &                          MCO \\
			67 &    \textbf{{\color{red}ENJ}} \\
			68 &                 \textbf{XMX} \\
			69 &                         ARDR \\
			70 &   \textbf{{\color{red}NEXO}} \\
			71 &                 \textbf{ZIL} \\
			72 &                         CRPT \\
			73 &                         NPXS \\
			74 &    \textbf{{\color{red}GNT}} \\
			75 &  \textbf{{\color{red}STEEM}} \\
			76 &    \textbf{{\color{red}SNT}} \\
			77 &    \textbf{{\color{red}XZC}} \\
			78 &                 \textbf{REN} \\
			79 &                         BTMX \\
			80 &                          YOU \\
			81 &   \textbf{{\color{red}MANA}} \\
			82 &                          ELF \\
			83 &    \textbf{{\color{red}ETN}} \\
			84 &                           BZ \\
			85 &                          MOF \\
			86 &  \textbf{{\color{red}STRAT}} \\
			87 &                          AOA \\
			88 &                           ZB \\
			89 &                 \textbf{ETP} \\
			90 &                 \textbf{KNC} \\
			91 &                          LRC \\
			92 &    \textbf{{\color{red}FCT}} \\
			93 &  \textbf{{\color{red}MATIC}} \\
			94 &                 \textbf{RDD} \\
			95 &    \textbf{{\color{red}ZEN}} \\
			96 &    \textbf{{\color{red}TNT}} \\
			97 &    \textbf{{\color{red}ELA}} \\
			98 &                \textbf{BEAM} \\
			99 &                         HBAR \\
			100 &    \textbf{{\color{red}ARK}} \\
			101 &               \textbf{SOLVE} \\
			102 &               \textbf{ERC20} \\
			103 &   \textbf{{\color{red}GRIN}} \\
			104 &                \textbf{NULS} \\
			105 &                          DGD \\
			106 &   \textbf{{\color{red}AION}} \\
			107 &                          YCC \\
			108 &    \textbf{{\color{red}ENG}} \\
			109 &                 \textbf{RLC} \\
			110 &                          FTM \\
			111 &                          ANT \\
			112 &                          CHZ \\
			113 &                         QASH \\
			114 &                   \textbf{R} \\
			115 &    \textbf{{\color{red}BNT}} \\
			116 &    \textbf{{\color{red}NAS}} \\
			117 &    \textbf{{\color{red}WAN}} \\
			118 &   \textbf{{\color{red}TOMO}} \\
			119 &    \textbf{{\color{red}NMR}} \\
			120 &                          WTC \\
			121 &                 \textbf{RCN} \\
			122 &                         POWR \\
			123 &                  \textbf{TT} \\
			124 &                 \textbf{QKC} \\
			125 &   \textbf{{\color{red}IOTX}} \\
			126 &                         DENT \\
			127 &  \textbf{{\color{red}STORJ}} \\
			128 &                          MTL \\
			129 &                          GAS \\
			130 &                        IGNIS \\
			\bottomrule
		\end{tabular}
		\begin{tabular}{lr}
			\toprule
			rank &      crypto\\
			\midrule
			131 &                          BRD \\
			132 &                          BIX \\
			133 &                 \textbf{PPT} \\
			134 &    \textbf{{\color{red}MDA}} \\
			135 &                          GNO \\
			136 &                \textbf{CELR} \\
			137 &                 \textbf{ONE} \\
			138 &    \textbf{{\color{red}GRS}} \\
			139 &   \textbf{{\color{red}PIVX}} \\
			140 &                        COCOS \\
			141 &                          ABT \\
			142 &  \textbf{{\color{red}GBYTE}} \\
			143 &   \textbf{{\color{red}LOOM}} \\
			144 &    \textbf{{\color{red}SYS}} \\
			145 &                 \textbf{RSR} \\
			146 &                 \textbf{NXS} \\
			147 &   \textbf{{\color{red}POLY}} \\
			148 &                          KAN \\
			149 &                          CND \\
			150 &                 \textbf{FET} \\
			151 &                \textbf{EURS} \\
			152 &                          HYN \\
			153 &                          NXT \\
			154 &                 \textbf{APL} \\
			155 &                          TKN \\
			156 &                 \textbf{DCN} \\
			157 &                          UIP \\
			158 &                          WXT \\
			159 &    \textbf{{\color{red}VTC}} \\
			160 &                          EDO \\
			161 &                         DUSK \\
			162 &                          SKY \\
			163 &    \textbf{{\color{red}MAN}} \\
			164 &                 \textbf{FSN} \\
			165 &   \textbf{{\color{red}EMC2}} \\
			166 &                 \textbf{HYC} \\
			167 &   \textbf{{\color{red}DRGN}} \\
			168 &                 \textbf{REQ} \\
			169 &                          UTK \\
			170 &                \textbf{ANKR} \\
			171 &    \textbf{{\color{red}QRL}} \\
			172 &                          GTO \\
			173 &                         CVNT \\
			174 &   \textbf{{\color{red}PART}} \\
			175 &                          PPP \\
			176 &                         LEND \\
			177 &                \textbf{DATA} \\
			178 &                 \textbf{ITC} \\
			179 &                         PERL \\
			180 &    \textbf{{\color{red}MFT}} \\
			181 &                 \textbf{INT} \\
			182 &                 \textbf{OCN} \\
			183 &               \textbf{STORM} \\
			184 &                          CDT \\
			185 &   \textbf{{\color{red}IRIS}} \\
			186 &                 \textbf{TNB} \\
			187 &    \textbf{{\color{red}OST}} \\
			188 &                          1ST \\
			189 &                         WABI \\
			190 &                \textbf{RUFF} \\
			191 &                           GO \\
			192 &                          EGT \\
			193 &    \textbf{{\color{red}KEY}} \\
			194 &                          EVX \\
			195 &               \textbf{BURST} \\
			\bottomrule
		\end{tabular}
		\begin{tabular}{lr}
			\toprule
			rank &      crypto\\
			\midrule
			196 &                 \textbf{INS} \\
			197 &                 \textbf{LBA} \\
			198 &                \textbf{MITH} \\
			199 &                          ADX \\
			200 &                \textbf{NEBL} \\
			201 &    \textbf{{\color{red}LOC}} \\
			202 &    \textbf{{\color{red}PPC}} \\
			203 &                          BAY \\
			204 &                         SALT \\
			205 &                          HPB \\
			206 &  \textbf{{\color{red}BLOCK}} \\
			207 &                          NPX \\
			208 &    \textbf{{\color{red}POE}} \\
			209 &                \textbf{DOCK} \\
			210 &                 \textbf{XAS} \\
			211 &                          PRO \\
			212 &                 \textbf{FLO} \\
			213 &                         BAND \\
			214 &    \textbf{{\color{red}NAV}} \\
			215 &    \textbf{{\color{red}SMT}} \\
			216 &                        SNGLS \\
			217 &                \textbf{TRIO} \\
			218 &                          SOC \\
			219 &   \textbf{{\color{red}TRUE}} \\
			220 &                          DBC \\
			221 &               \textbf{AERGO} \\
			222 &    \textbf{{\color{red}SNM}} \\
			223 &                 \textbf{MTH} \\
			224 &                          SBD \\
			225 &                 \textbf{DNT} \\
			226 &                 \textbf{GVT} \\
			227 &    \textbf{{\color{red}VIA}} \\
			228 &                 \textbf{VEE} \\
			229 &                 \textbf{UCT} \\
			230 &    \textbf{{\color{red}WPR}} \\
			231 &                         STPT \\
			232 &    \textbf{{\color{red}AST}} \\
			233 &                \textbf{CONI} \\
			234 &                 \textbf{XDN} \\
			235 &               \textbf{SWFTC} \\
			236 &                         AUTO \\
			237 &                 \textbf{SRN} \\
			238 &    \textbf{{\color{red}AMB}} \\
			239 &                          VIB \\
			240 &                          ENQ \\
			241 &                          QUN \\
			242 &                 \textbf{QLC} \\
			243 &                 \textbf{MDS} \\
			244 &                 \textbf{DDD} \\
			245 &                         DROP \\
			246 &                  \textbf{LA} \\
			247 &                          BTX \\
			248 &                         APPC \\
			249 &                          ZIP \\
			250 &   \textbf{{\color{red}CLAM}} \\
			251 &    \textbf{{\color{red}XCP}} \\
			252 &    \textbf{{\color{red}POA}} \\
			253 &                \textbf{VIBE} \\
			254 &                 \textbf{ARN} \\
			255 &                          LYM \\
			256 &                 \textbf{VBK} \\
			257 &                          PST \\
			258 &                          OAX \\
			259 &    \textbf{{\color{red}UBQ}} \\
			260 &                \textbf{DADI} \\
			\bottomrule
		\end{tabular}
		\begin{tabular}{lr}
			\toprule
			rank &      crypto\\
			\midrule
			261 &   \textbf{{\color{red}FUEL}} \\
			262 &                 \textbf{DTA} \\
			263 &                         BCPT \\
			264 &                 \textbf{DLT} \\
			265 &    \textbf{{\color{red}EMC}} \\
			266 &                          VID \\
			267 &                          BLK \\
			268 &                 \textbf{GNX} \\
			269 &                 \textbf{YEE} \\
			270 &                          OGO \\
			271 &                          BMX \\
			272 &                          BTO \\
			273 &    \textbf{{\color{red}MLN}} \\
			274 &                 \textbf{PNT} \\
			275 &                 \textbf{LND} \\
			276 &                          UPP \\
			277 &                          CHR \\
			278 &                          PRA \\
			279 &                        WINGS \\
			280 &    \textbf{{\color{red}DPY}} \\
			281 &               \textbf{NCASH} \\
			282 &                 \textbf{PCH} \\
			283 &                 \textbf{LUN} \\
			284 &                 \textbf{SNC} \\
			285 &                 \textbf{OLT} \\
			286 &                          SKM \\
			287 &                          ADT \\
			288 &                        YOYOW \\
			289 &    \textbf{{\color{red}XPM}} \\
			290 &                  \textbf{OK} \\
			291 &                \textbf{KICK} \\
			292 &    \textbf{{\color{red}FTC}} \\
			293 &                         BZNT \\
			294 &    \textbf{{\color{red}LBC}} \\
			295 &                 \textbf{CZR} \\
			296 &                          ATN \\
			297 &    \textbf{{\color{red}NTY}} \\
			298 &                          NGC \\
			299 &                          KIN \\
			300 &                          DPN \\
			301 &                 \textbf{SUB} \\
			302 &                          IHT \\
			303 &               \textbf{CLOAK} \\
			304 &                 \textbf{REM} \\
			305 &    \textbf{{\color{red}INK}} \\
			306 &                          SSP \\
			307 &                 \textbf{DCT} \\
			308 &                 \textbf{EDG} \\
			309 &                          ZCL \\
			310 &                 \textbf{PHR} \\
			311 &   \textbf{{\color{red}GAME}} \\
			312 &                \textbf{PASC} \\
			313 &                 \textbf{ZPT} \\
			314 &    \textbf{{\color{red}LUX}} \\
			315 &                         CHAT \\
			316 &                 \textbf{PTT} \\
			317 &    \textbf{{\color{red}XMY}} \\
			318 &                          EXP \\
			319 &    \textbf{{\color{red}BOX}} \\
			320 &                \textbf{RADS} \\
			321 &    \textbf{{\color{red}POT}} \\
			322 &                \textbf{UBEX} \\
			323 &                \textbf{APIS} \\
			324 &                        AIDOC \\
			325 &                 \textbf{EKO} \\
			\bottomrule
		\end{tabular}
		\begin{tabular}{lr}
			\toprule
			rank &      crypto\\
			\midrule
			326 &                 \textbf{XHV} \\
			327 &                          EDU \\
			328 &                 \textbf{VRC} \\
			329 &                          DAX \\
			330 &                         SENC \\
			331 &                 \textbf{AAC} \\
			332 &                 \textbf{TOS} \\
			333 &                          MGO \\
			334 &   \textbf{{\color{red}VITE}} \\
			335 &   \textbf{{\color{red}TIME}} \\
			336 &                          BGG \\
			337 &    \textbf{{\color{red}BKX}} \\
			338 &                 \textbf{SSC} \\
			339 &                          DAT \\
			340 &    \textbf{{\color{red}HMQ}} \\
			341 &                         COTI \\
			342 &   \textbf{{\color{red}MYST}} \\
			343 &    \textbf{{\color{red}SWT}} \\
			344 &               \textbf{SHIFT} \\
			345 &                 \textbf{ION} \\
			346 &                 \textbf{FXT} \\
			347 &                 \textbf{EFX} \\
			348 &                 \textbf{SIB} \\
			349 &                        COVAL \\
			350 &                          MUE \\
			351 &   \textbf{{\color{red}OMNI}} \\
			352 &                          RVR \\
			353 &                \textbf{COVA} \\
			354 &                 \textbf{XEL} \\
			355 &                \textbf{CAPP} \\
			356 &                          LEO \\
			357 &                 \textbf{MTX} \\
			358 &                         BLOC \\
			359 &                        SPHTX \\
			360 &                          IXT \\
			361 &                          UQC \\
			362 &    \textbf{{\color{red}XBC}} \\
			363 &                \textbf{PINK} \\
			364 &                         OPEN \\
			365 &                \textbf{MNTP} \\
			366 &                          THC \\
			367 &   \textbf{{\color{red}EXCL}} \\
			368 &                \textbf{KORE} \\
			369 &                         WICC \\
			370 &                        BANCA \\
			371 &    \textbf{{\color{red}AMP}} \\
			372 &                        MUSIC \\
			373 &                          MER \\
			374 &                         ORME \\
			375 &                \textbf{MEME} \\
			376 &                 \textbf{EXY} \\
			377 &                 \textbf{STQ} \\
			378 &                \textbf{SPHR} \\
			379 &                 \textbf{COB} \\
			380 &                 \textbf{MOT} \\
			381 &                          RCT \\
			382 &                 \textbf{GUP} \\
			383 &                         DOPE \\
			384 &                         NEOS \\
			385 &                          GEM \\
			386 &                         ARCT \\
			387 &                 \textbf{BDG} \\
			388 &                          ALI \\
			389 &                         BTCS \\
			390 &                 \textbf{BSD} \\
			\bottomrule
		\end{tabular}
		\begin{tabular}{lr}
			\toprule
			rank &      crypto\\
			\midrule
			391 &                         STAR \\
			392 &    \textbf{{\color{red}TIX}} \\
			393 &                 \textbf{BOT} \\
			394 &                          REX \\
			395 &                 \textbf{CPC} \\
			396 &                 \textbf{MIC} \\
			397 &                 \textbf{PFR} \\
			398 &                          PKB \\
			399 &                          DOR \\
			400 &                          LXT \\
			401 &     \textbf{{\color{red}TX}} \\
			402 &             \textbf{ELTCOIN} \\
			403 &                \textbf{SYNX} \\
			404 &                         EMPR \\
			405 &    \textbf{{\color{red}CHC}} \\
			406 &                \textbf{FTXT} \\
			407 &                          DAC \\
			408 &                \textbf{ETHM} \\
			409 &                  \textbf{HC} \\
			410 &                 \textbf{LMC} \\
			411 &                \textbf{ABBC} \\
			412 &                 \textbf{ADK} \\
			413 &                 \textbf{ADN} \\
			414 &                         AGRS \\
			415 &                 \textbf{B2X} \\
			416 &                          BBN \\
			417 &                          BCY \\
			418 &                         BEAN \\
			419 &                \textbf{BELA} \\
			420 &                      BITGOLD \\
			421 &                          BLZ \\
			422 &                 \textbf{BMC} \\
			423 &                          BT2 \\
			424 &                         BTCB \\
			425 &                \textbf{BTNT} \\
			426 &                \textbf{CACH} \\
			427 &   \textbf{{\color{red}CANN}} \\
			428 &                 \textbf{CCL} \\
			429 &                          CET \\
			430 &   \textbf{{\color{red}CLUB}} \\
			431 &                          CMT \\
			432 &                          CON \\
			433 &                \textbf{CTXC} \\
			434 &                          CVC \\
			435 &                 \textbf{CWV} \\
			436 &                        DAPPT \\
			437 &    \textbf{{\color{red}DAR}} \\
			438 &                         DARC \\
			439 &                          DEX \\
			440 &                \textbf{DIVX} \\
			441 &                         DOCT \\
			442 &                          DOT \\
			443 &                          DVP \\
			444 &                          DYN \\
			445 &   \textbf{{\color{red}EBST}} \\
			446 &                          EDT \\
			447 &                          EKT \\
			448 &                          ERD \\
			449 &                          ETF \\
			450 &                 \textbf{EVN} \\
			451 &                          EVR \\
			452 &                          FKX \\
			453 &                \textbf{FLDC} \\
			454 &                          FNB \\
			455 &                 \textbf{FTO} \\
			\bottomrule
		\end{tabular}
		\begin{tabular}{lr}
			\toprule
			rank &      crypto\\
			\midrule
			456 &                          GCR \\
			457 &                 \textbf{GTC} \\
			458 &                          HBT \\
			459 &                          HKG \\
			460 &                          HPT \\
			461 &                         HUSD \\
			462 &                 \textbf{INC} \\
			463 &    \textbf{{\color{red}IOP}} \\
			464 &                          IRA \\
			465 &                          JAR \\
			466 &                          KBR \\
			467 &                \textbf{LBTC} \\
			468 &                         MCAP \\
			469 &                 \textbf{MCC} \\
			470 &                         META \\
			471 &   \textbf{{\color{red}MINT}} \\
			472 &                          MOD \\
			473 &                 \textbf{MOL} \\
			474 &                          MYO \\
			475 &                          NBT \\
			476 &                          NKC \\
			477 &                         NOIA \\
			478 &                          NXC \\
			479 &                           OC \\
			480 &                  \textbf{OF} \\
			481 &                \textbf{ORBS} \\
			482 &                          PAI \\
			483 &                 \textbf{PAL} \\
			484 &                          PBT \\
			485 &                 \textbf{PCL} \\
			486 &                         PIXL \\
			487 &                          PLY \\
			488 &                         PURA \\
			489 &                 \textbf{QSP} \\
			490 &                        QWARK \\
			491 &                         REPO \\
			492 &                         RIPT \\
			493 &                         RISE \\
			494 &                       ROOBEE \\
			495 &                 \textbf{RRC} \\
			496 &                        SAFEX \\
			497 &                 \textbf{SAN} \\
			498 &                         SERV \\
			499 &                          SFU \\
			500 &                          SHR \\
			501 &                 \textbf{SLS} \\
			502 &                          STC \\
			503 &                          TAN \\
			504 &                          TAS \\
			505 &                          THR \\
			506 &                          THX \\
			507 &                         TOPC \\
			508 &                         TRIG \\
			509 &                 \textbf{TTT} \\
			510 &                \textbf{UBTC} \\
			511 &                         VEEN \\
			512 &                \textbf{VEST} \\
			513 &                 \textbf{WGP} \\
			514 &                 \textbf{WIN} \\
			515 &                          XBB \\
			516 &                          XBT \\
			517 &    \textbf{{\color{red}XBY}} \\
			518 &                 \textbf{XET} \\
			519 &                 \textbf{XIN} \\
			520-521 &         \textbf{XUC}-ZDR \\
			\bottomrule
		\end{tabular}
	}
\end{table}

\clearpage

%%%%%%%%%%%%%%%%%%%%%%%%%%%%%%%%%%%
%%%%%%%%%%% FIGURES %%%%%%%%%%%%%%%
%%%%%%%%%%%%%%%%%%%%%%%%%%%%%%%%%%%
\section{Figures}
%%% Fig. S1
\noindent {\bf Fig. S1.} \textbf{Event types and cryptocurrency developers community}. (A) The number of GitHub Push Event and accepted Pull Requests events (left panel) and the number of developers (right panel) for the top $20$ cryptocurrencies. The \emph{Ethereum} community (ETH) is the most active. (B) Developing activity by event types. The top panel shows the number of events per type, revealing ``PushEvent'' as the most frequent events. The bottom panel shows the number of developers per event type. ``WatchEvent'' is the event type triggered by more users, while ``PushEvents'' are triggered by less than one-fifth of the users.
\begin{figure}[h]
	\centering
	\includegraphics[width=1.01\linewidth]{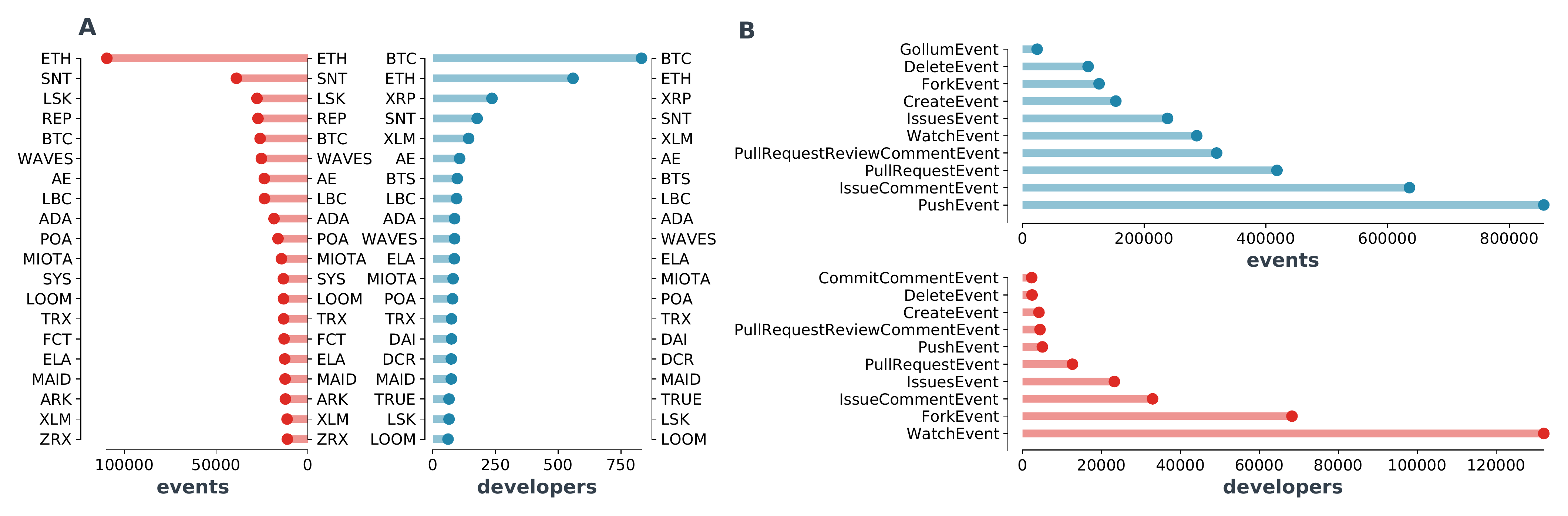}
	\label{SIfig1}
\end{figure}
\clearpage

%%% Fig. S2
\noindent {\bf Fig. S2.} \textbf{Developers and market / coding activity}. (A) Market capitalization vs the number of project developers for individual currencies (blue dots, Spearman correlation=0.49***) and groups of currencies with the same number of developers (red dots, Spearman correlation=0.72***). In the latter case, the y-axis reports the median market capitalization. (B) The number of edits vs the number of developers for individual currencies (blue dots, Spearman correlation=0.92***) and groups of currencies with the same number of developers (red dots, Spearman correlation=0.92***).
\begin{figure}[h]
	\centering
	\includegraphics[width=0.99\linewidth]{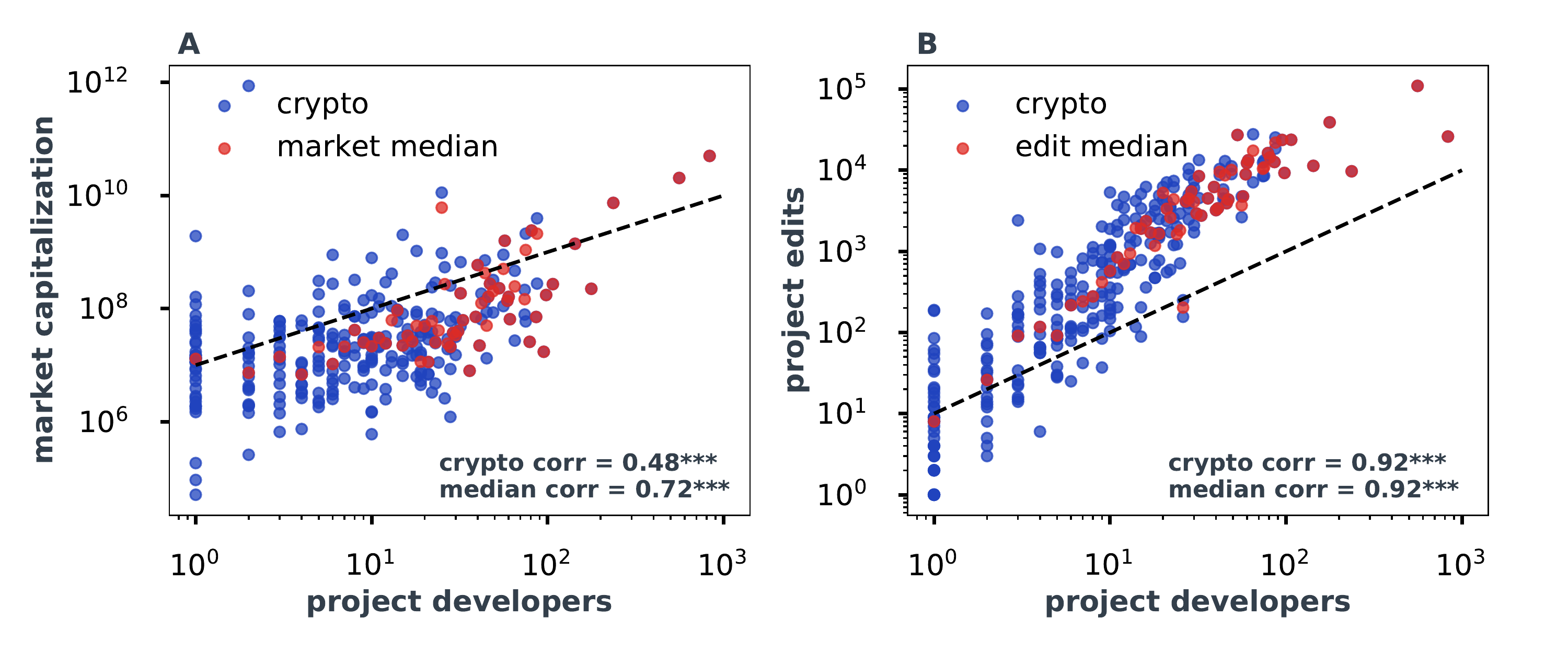}
	\label{SIfig2}
\end{figure}
\clearpage

%%% Fig. S3
\noindent {\bf Fig. S3.} \add{\textbf{Number of connections per developer}. Figure shows the distribution of the number of connections for which each developer connecting a pairs is responsible for. The inset graphs shows the community of linked cryptocurrencies for the three developers connecting more projects.}
\begin{figure}[h]
	\centering
	\includegraphics[width=0.99\linewidth]{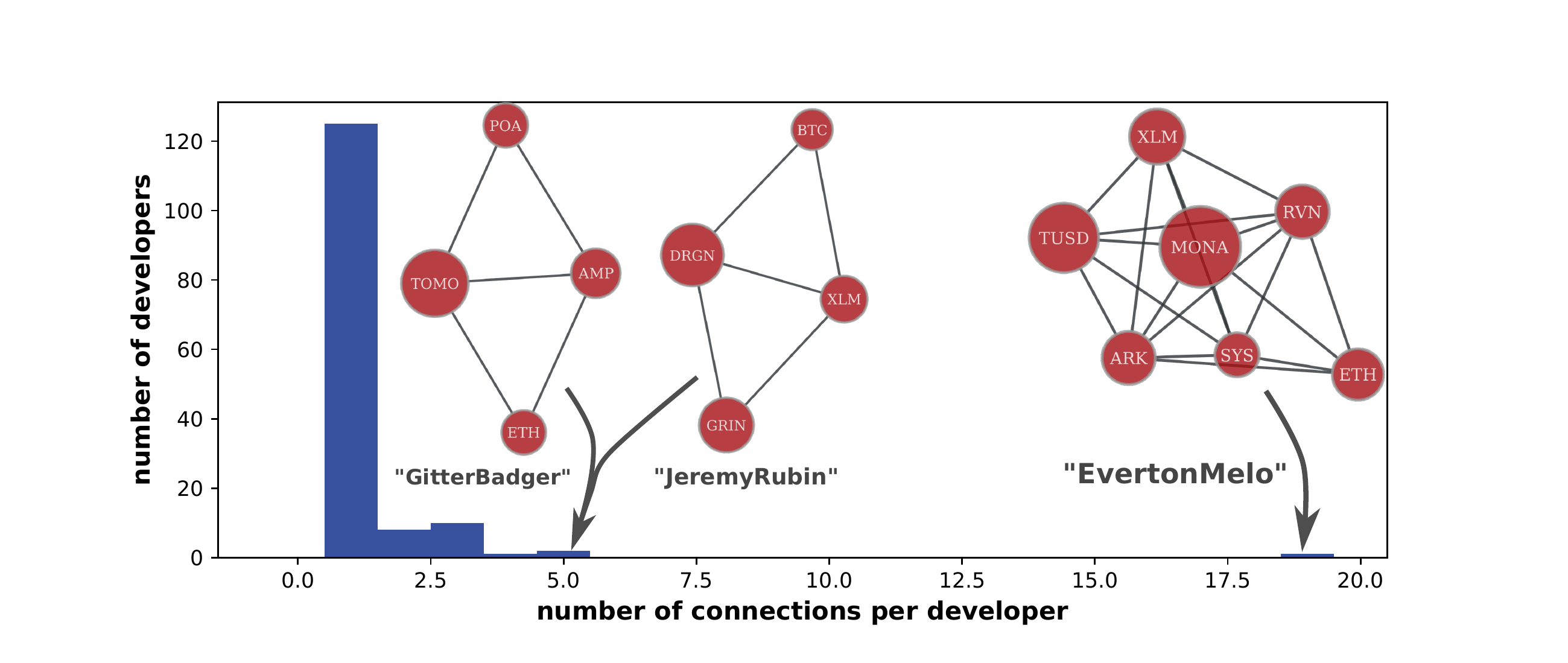}
	\label{SIfig3}
\end{figure}
\clearpage

%%% Fig. S4
\noindent {\bf Fig. S4.} \textbf{Degree distribution and link creation over time}. (A) The degree distribution of the network at the beginning of June 2019 (circles). (B) The number of new edges as a function of time (circles). The blue line captures the exponential growth of the new collaborations (doubling every $\sim 15$ months), where $x_d$ is the number of days from the first coupling event. Shaded areas represent the standard deviation of the models estimated from the MCMC fit.
\begin{figure}[h]
	\centering
	\includegraphics[width=0.99\linewidth]{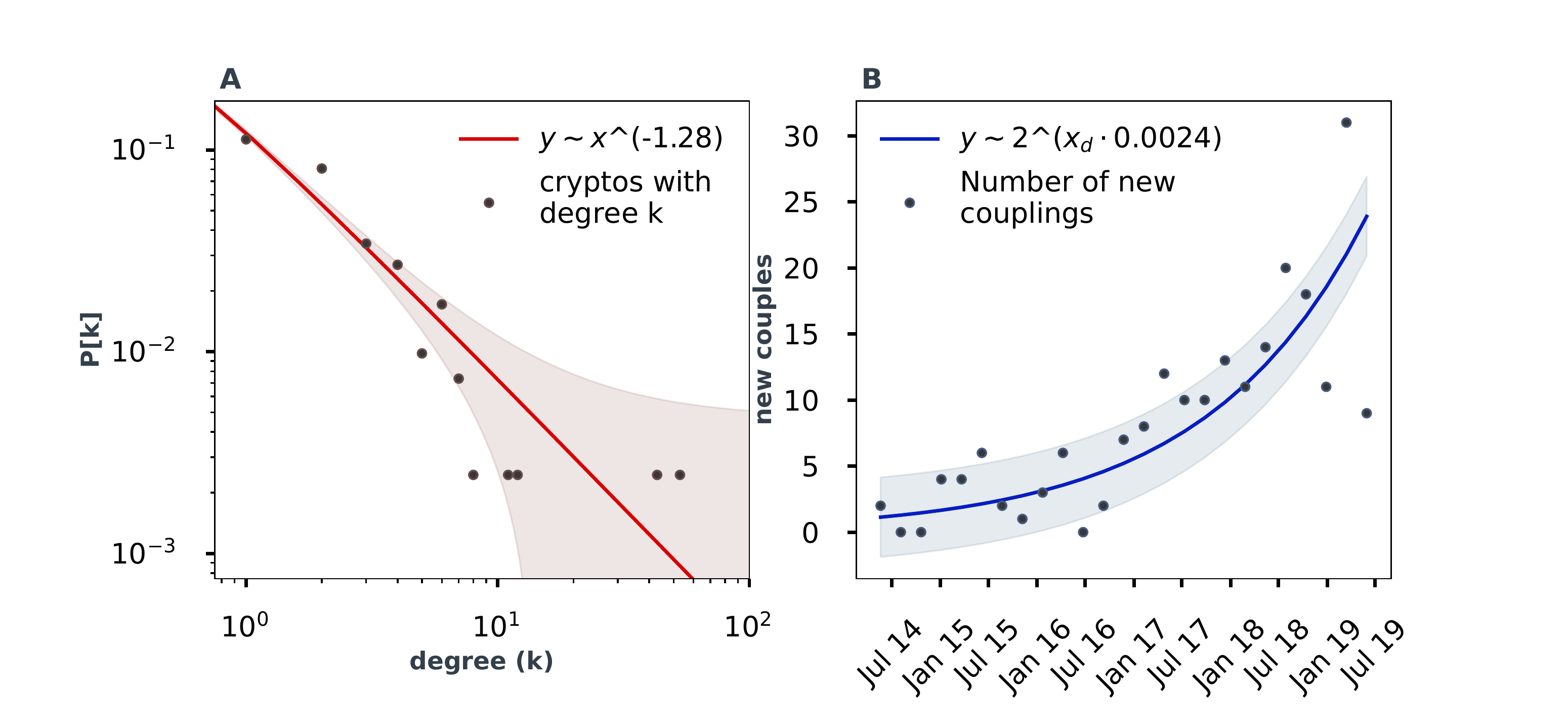}
	\label{SIfig4}
\end{figure}
\clearpage

%%% Fig. S5
\noindent {\bf Fig. S5.} \textbf{``PushEvent'' vs ``PullRequestEvent'' links over time}. (A) The number of new connections originated by a ``PushEvent'' (red bars) or a ``PullRequestEvent'' over time. The number of contributions due to accepted ``PullRequestEvents'' has grown over time. (B) The number of new connections in time by connection direction: the currency with highest capitalization is the first-edited (red bars) or second edited (blue bars). In yellow we report the number of pairs for which at least one component has null market capitalization before the connection time.
\begin{figure}[h]
	\centering
	\includegraphics[width=0.9\linewidth]{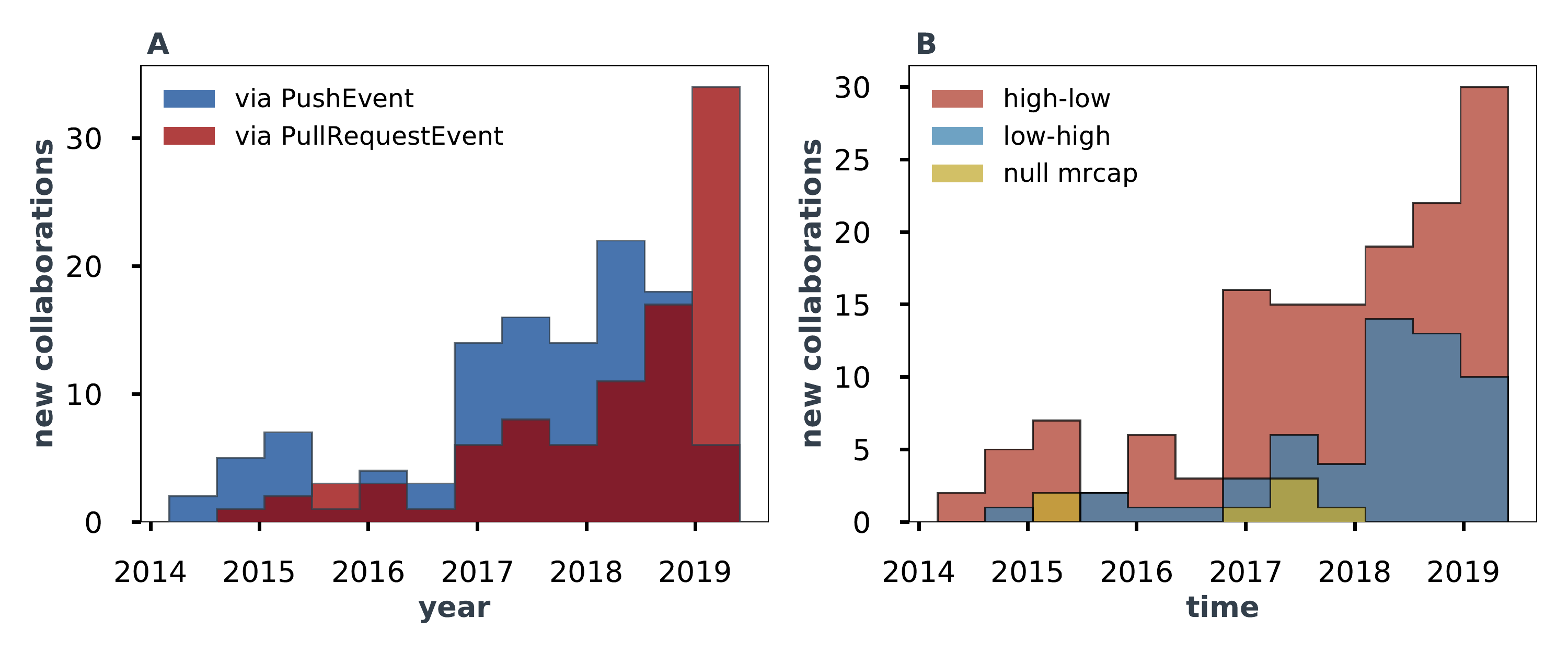}
	\label{SIfig5}
\end{figure}
\clearpage

%%% Fig. S6
\noindent {\bf Fig. S6.} \textbf{Robustness with respect to window sizes}. The standardised Spearman correlation (A) to (F) under different window sizes. Shaded areas represent $2$ standard deviations of the averages and are determined via bootstrap procedure (see Sec. Methods).
\begin{figure}[h]
	\centering
	\includegraphics[width=0.9\linewidth]{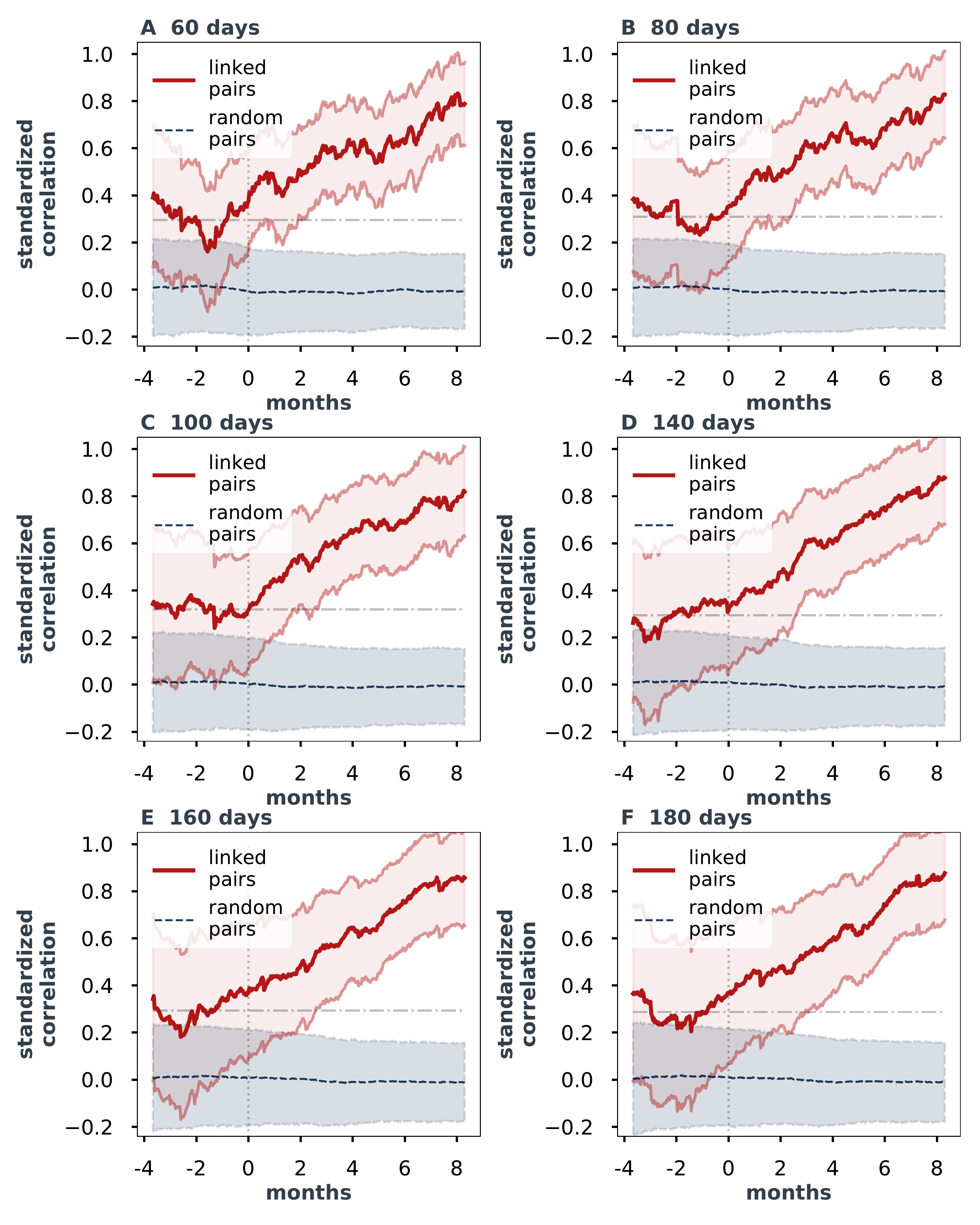}
	\label{SIfig6}
\end{figure}
\clearpage

%%% Fig. S7
\noindent {\bf Fig. S7.} \textbf{Increase in correlation under different window sizes}. (A-I) The distribution of differences in standardised correlation between after and before connection time. Results are shown for different window sizes, reported above each graph.
\begin{figure}[t]
	\centering
	\includegraphics[width=0.8\paperwidth]{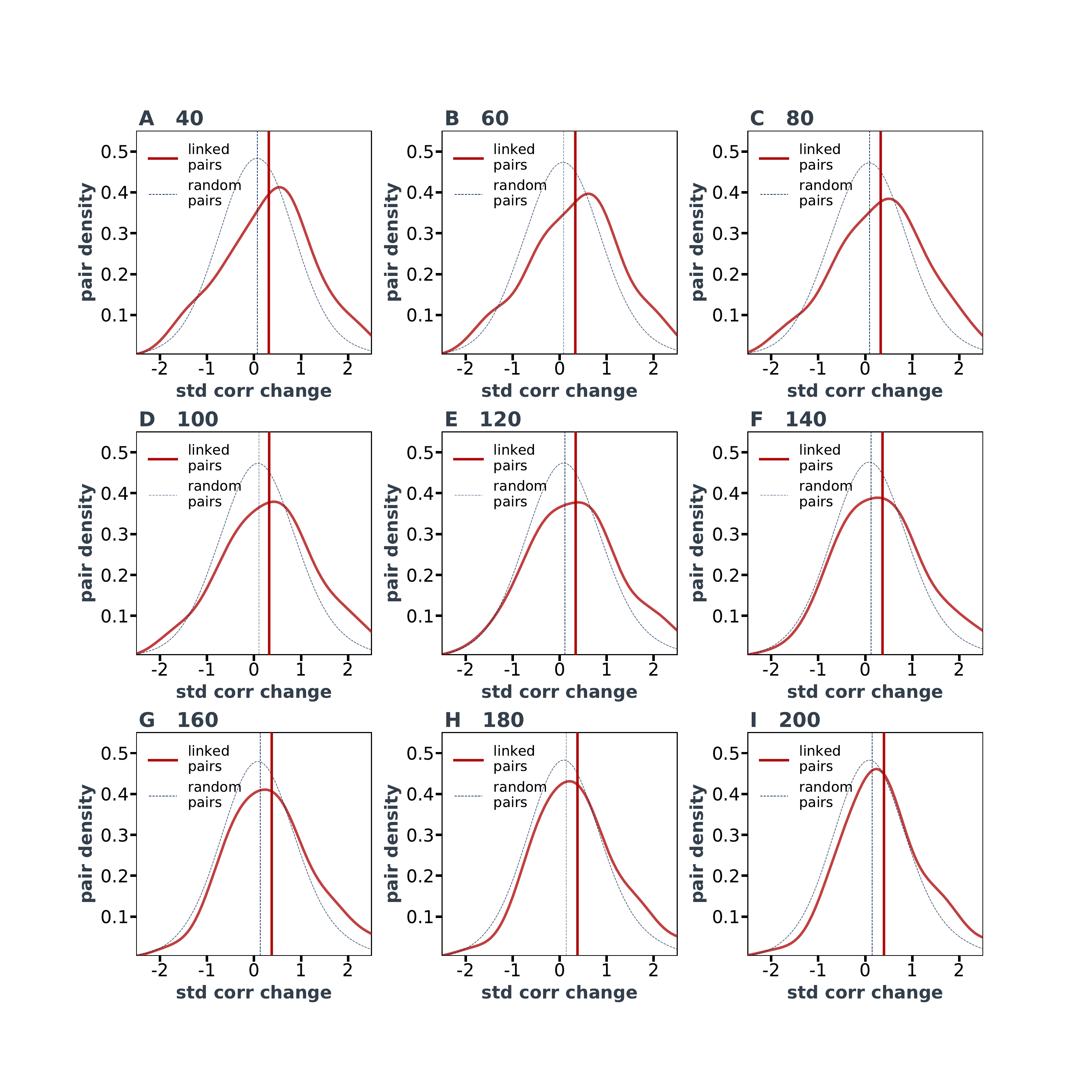}
	\label{SIfig7}
\end{figure}
\clearpage

%%% Fig. S8
\noindent {\bf Fig. S8.} \textbf{Results are robust across various randomizations}. (A) The Spearman correlation for linked pairs (red continuous line), RT pairs (blue dotted line), RTA pairs (green dot-dashed line) and ORTA pairs (purple dashed line). (B) The standardised correlation with respect to the average of the entire ecology. Shaded areas represent $2$ standard deviations of the averages and are determined via bootstrap procedure (see Sec. Methods).
\begin{figure}[h]
	\centering
	\includegraphics[width=0.99\linewidth]{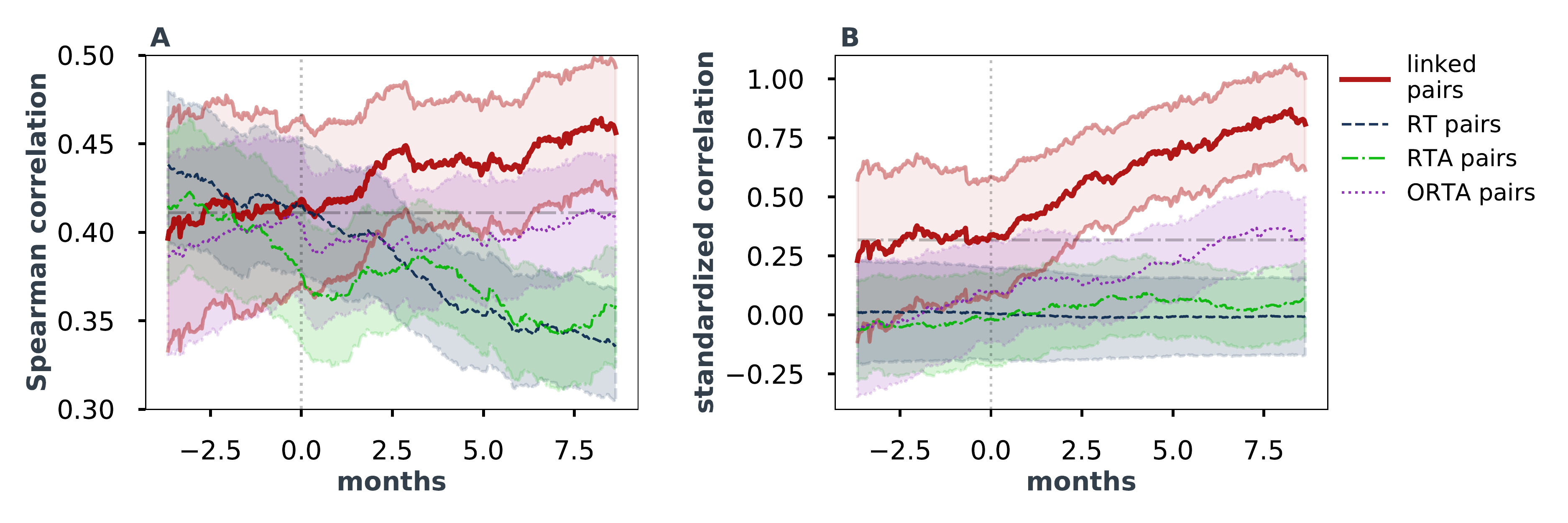}
	\label{SIfig8}
\end{figure}
\clearpage

%%% Fig. S9
\noindent {\bf Fig. S9.} \textbf{Results are robust across linked pairs subsampling}. (A) The standardised Spearman correlation for linked pairs without BTC (orange continuous line) and RT pairs (blue dotted line). (B) The standardised Spearman correlation for linked pairs without ETH (orange continuous line) and RT pairs (blue dotted line). Shaded areas represent $2$ standard deviations of the averages and are determined via bootstrap procedure (see Sec. Methods).
\begin{figure}[h]
	\centering
	\includegraphics[width=0.99\linewidth]{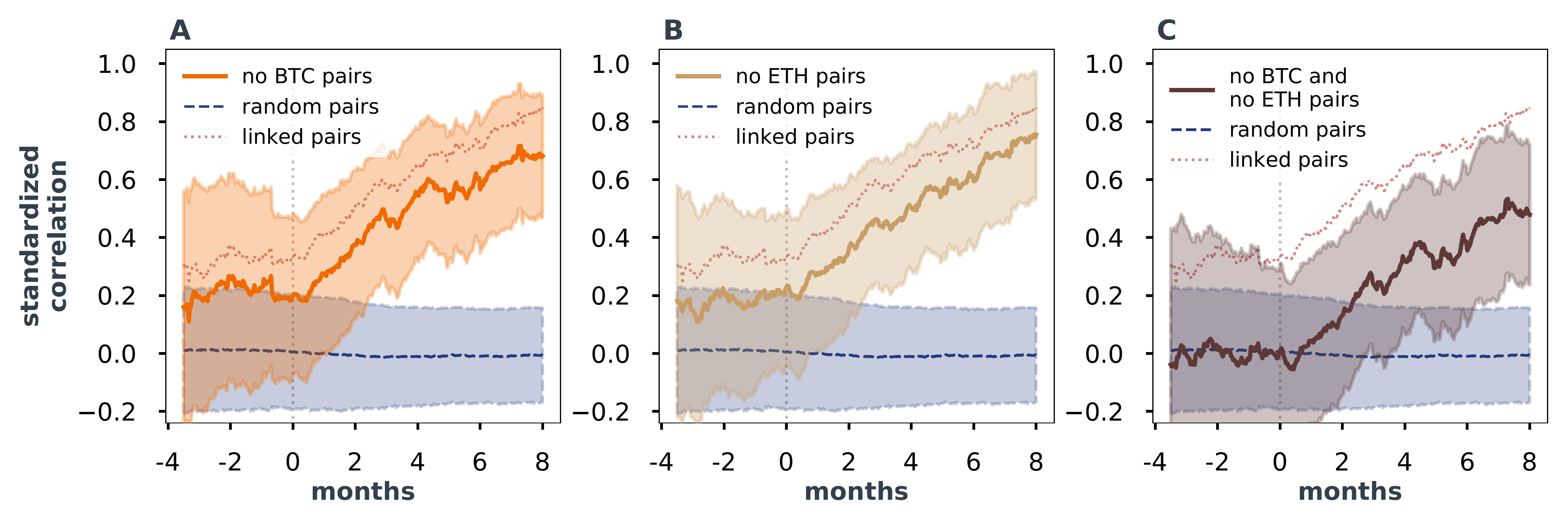}
	\label{SIfig9}
\end{figure}
\clearpage

%%% Fig. S10
\noindent {\bf Fig. S10.} \textbf{Results are robust across connection-intensity subsampling of linked pairs}. The standardised Spearman correlation for linked pairs with only one link. Pairs which in the time after the connections incurred by subsequent link activations by different developers are dropped in the ``one-link'' configuration. Shaded areas represent $2$ standard deviations of the averages and are determined via bootstrap procedure (see Sec. Methods).
\begin{figure}[h]
	\centering
	\includegraphics[width=0.5    \linewidth]{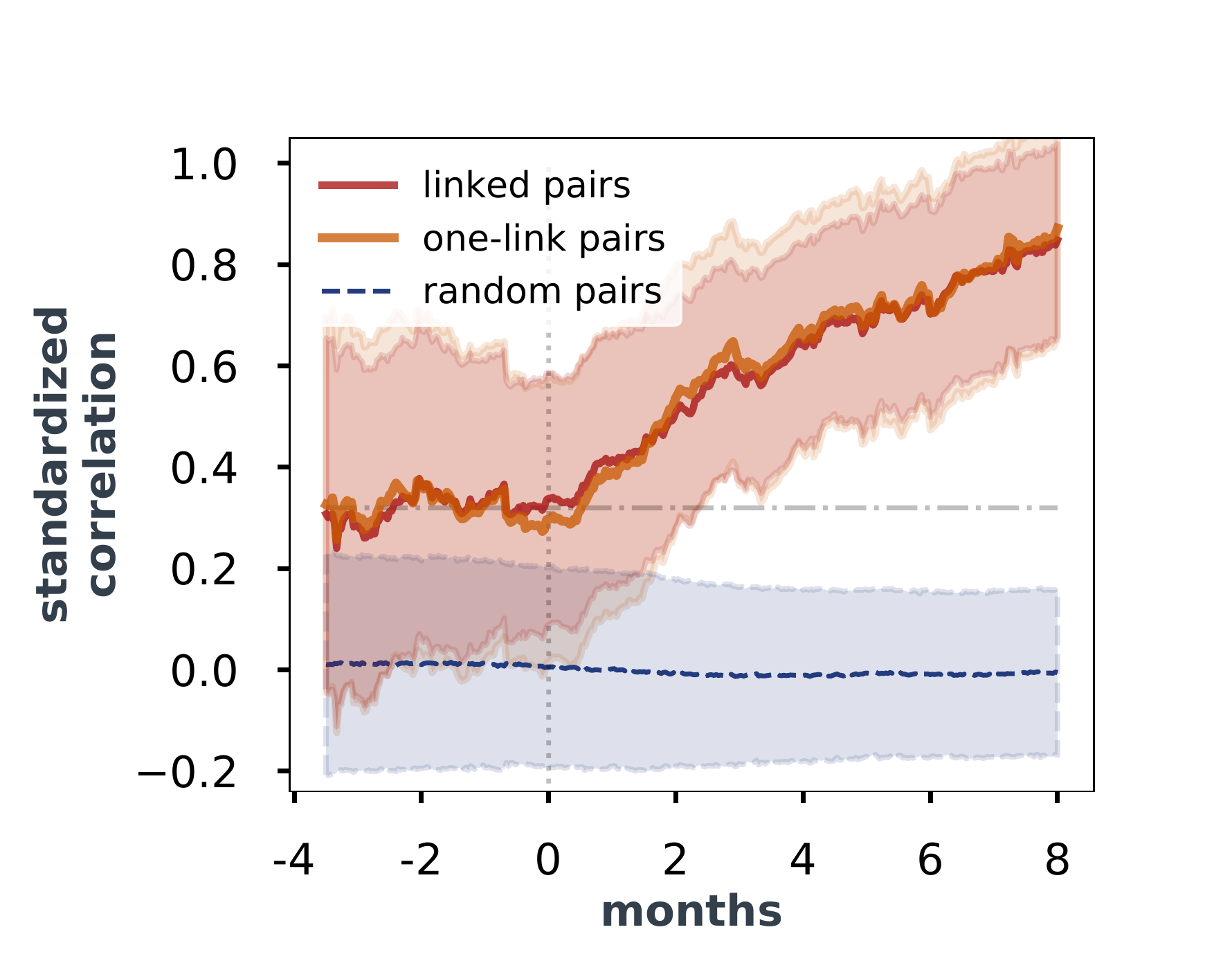}
	\label{SIfig10}
\end{figure}
\clearpage

%%% Fig. S11
\noindent {\bf Fig. S11.} \textbf{Synchronization of the median standardised correlation}. Continuous red line report the median of linked pairs (bootstrapped following the procedure presented in Materials and Methods). Dashed blue line shows the median value of the standardised correlation for the random pairs. Shaded areas represent the $[0.025,.0975]$ quantile region for the median.
\begin{figure}[h]
	\centering
	\includegraphics[width=0.5    \linewidth]{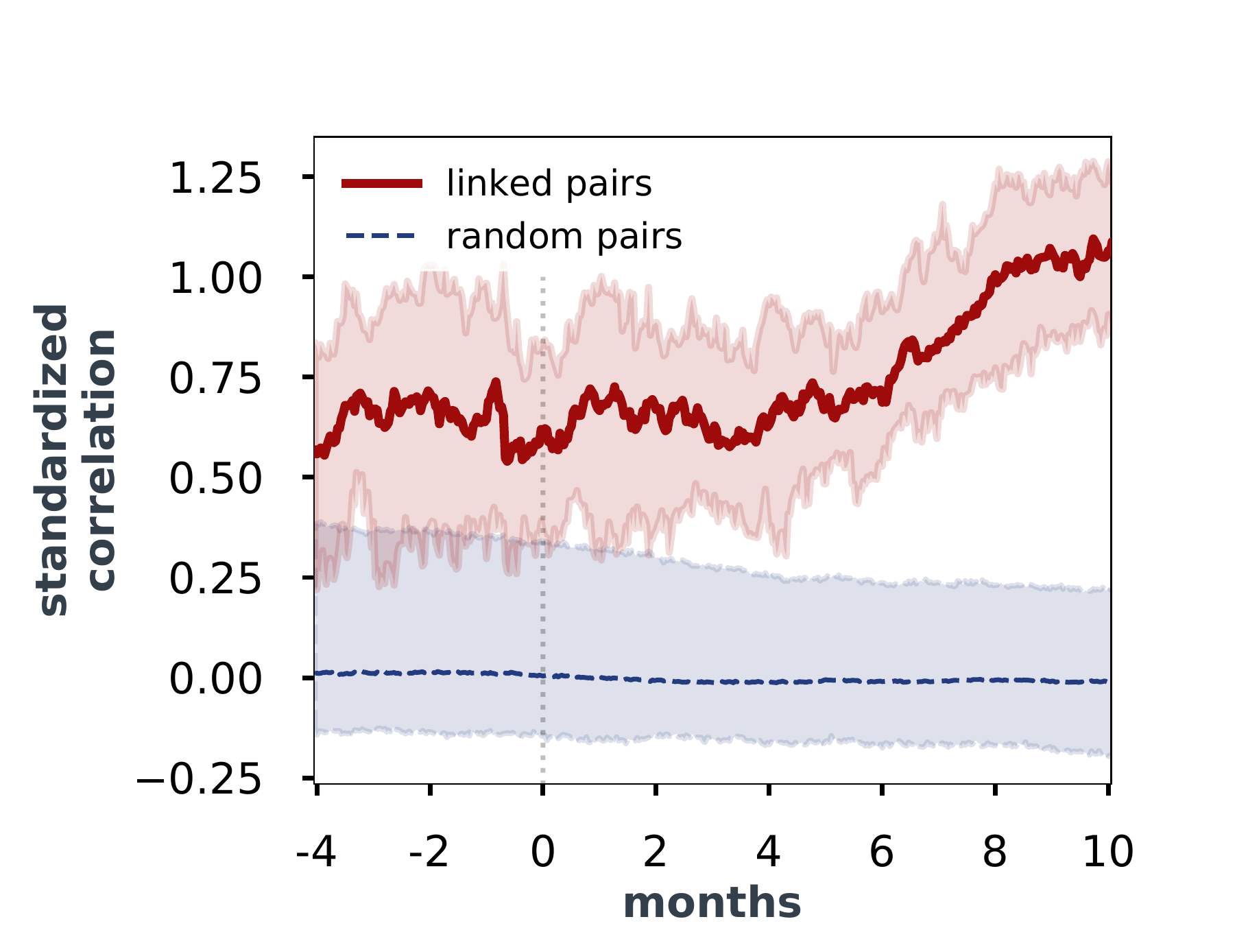}
	\label{SIfig11}
\end{figure}
\clearpage

%%% Fig. S12
\noindent {\bf Fig. S12.} \textbf{Raw histograms of correlation change}. Full-red histogram reports the raw distribution of the standardised correlation change for linked pairs. Dashed-blue histogram reports the raw distribution of the standardised correlation change for random pairs (RT pairs). Vertical lines highlight the average of the distributions. In particular, vertical continuous red line shows the average of the linked pair distribution while the dashed-blue one shows the average of the random pair distribution.
\begin{figure}[h]
	\centering
	\includegraphics[width=0.5    \linewidth]{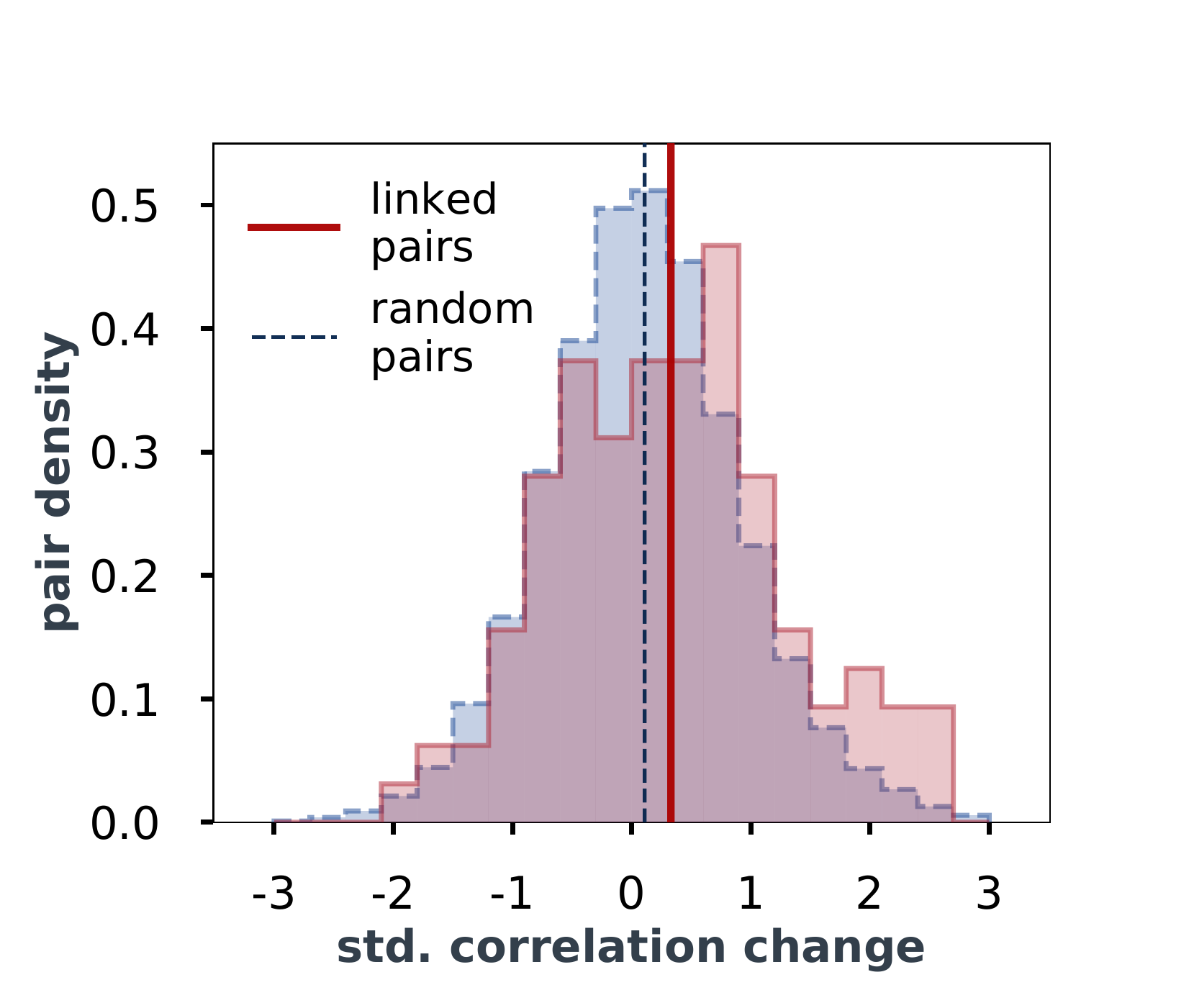}
	\label{SIfig12}
\end{figure}
\clearpage

%%% Fig. S13
\noindent {\bf Fig. S13.} \textbf{Fraction of increasing standardised correlation}. Continuous red line shows the average fraction of linked pairs whose standardised correlation increases, on average, after the connection time on a window of size x. Dashed blue line shows the fraction of random pairs whose standardised correlation is increasing. Shaded areas represent $2$ standard deviations of the averages and are determined via bootstrap procedure (see Sec. Methods).
\begin{figure}[h]
	\centering
	\includegraphics[width=0.5    \linewidth]{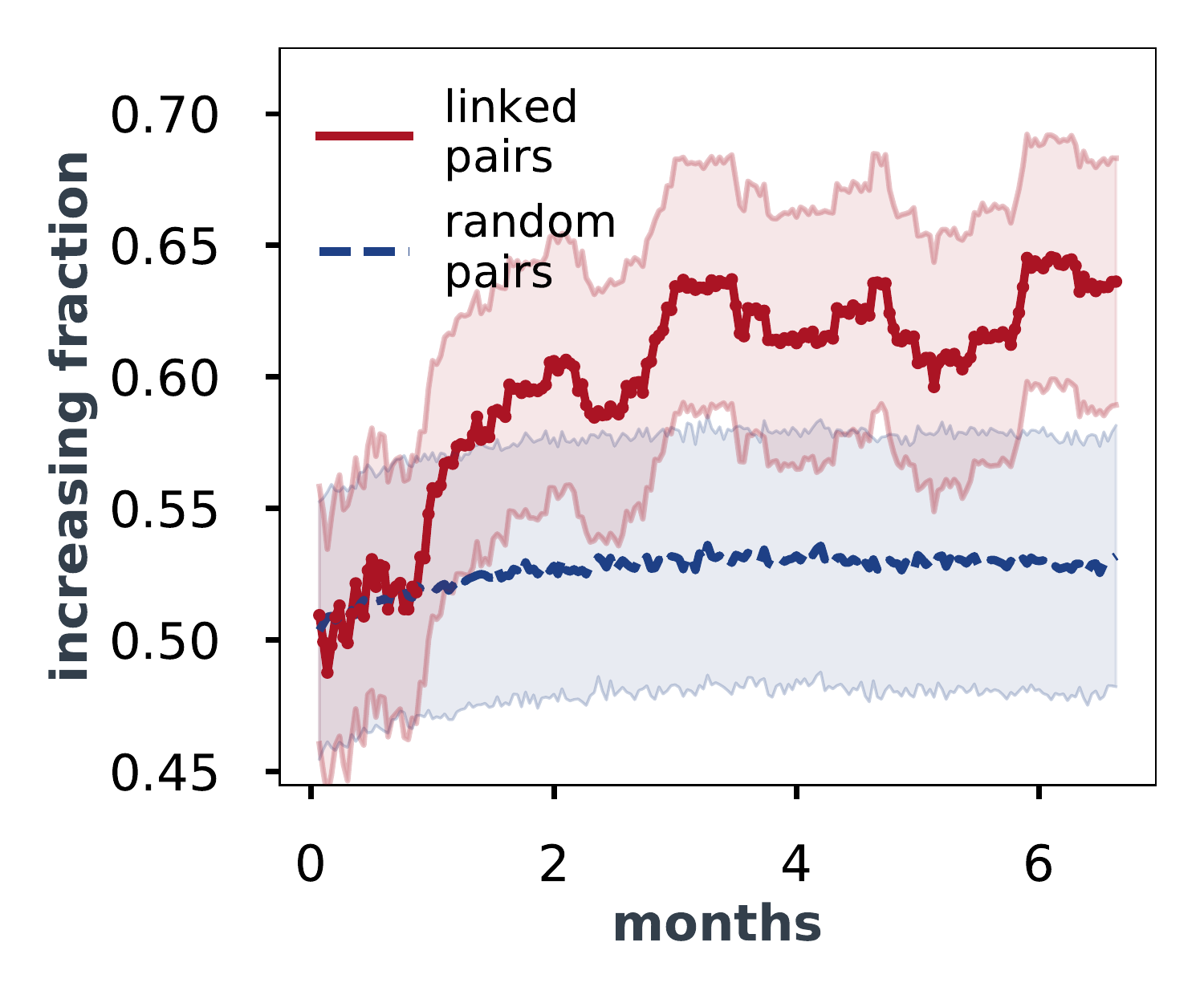}
	\label{SIfig13}
\end{figure}
\clearpage

%%% Fig. S14
\noindent {\bf Fig. S14.} \textbf{Correlation of various market metrics centred at connection time}. (A-F) The standardised correlations, centred at the connection time, for linked pairs, RT pairs, RTA pairs, and ORTA pairs. The measures studied are: transaction volume (A), volume change (B), market capitalisation (C), capitalisation change (D), price (E), and volatility (F). Shaded areas correspond to $2$ standard deviation intervals.
\begin{figure}[t]
	\centering
	\includegraphics[width=0.99\linewidth]{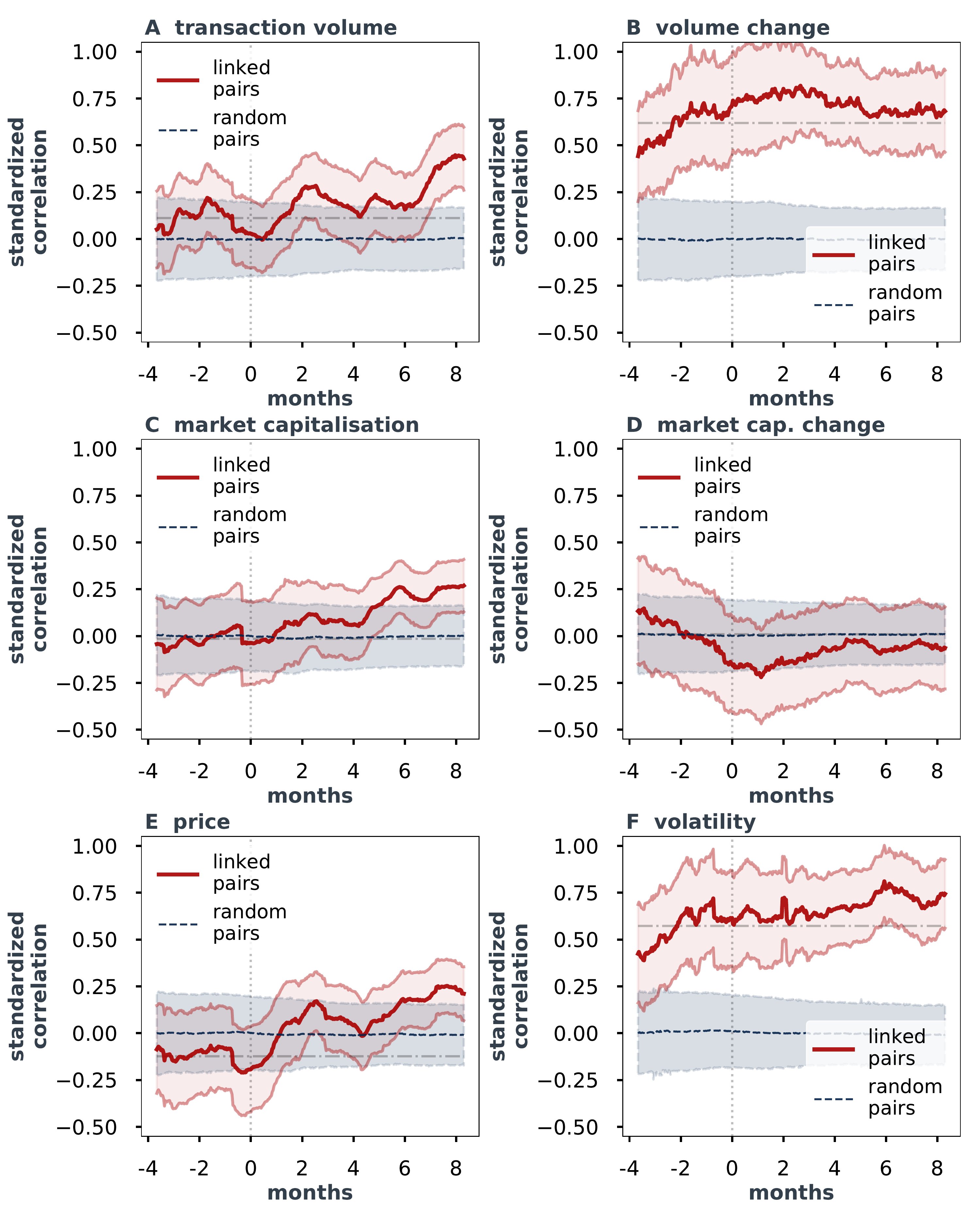}
	\label{SIfig14}
\end{figure}
\clearpage

%%% Fig. S15
\noindent {\bf Fig. S15.} \textbf{Cryptocurrencies in a linked pair differ in terms of age.}. Distribution of market age for linked cryptocurrencies (A) and random cryptocurrencies (B). Full line correspond to the distribution of the oldest crypto in the pair, dashed line to the youngest.
\begin{figure}[h]
	\centering
	\includegraphics[width=0.99\linewidth]{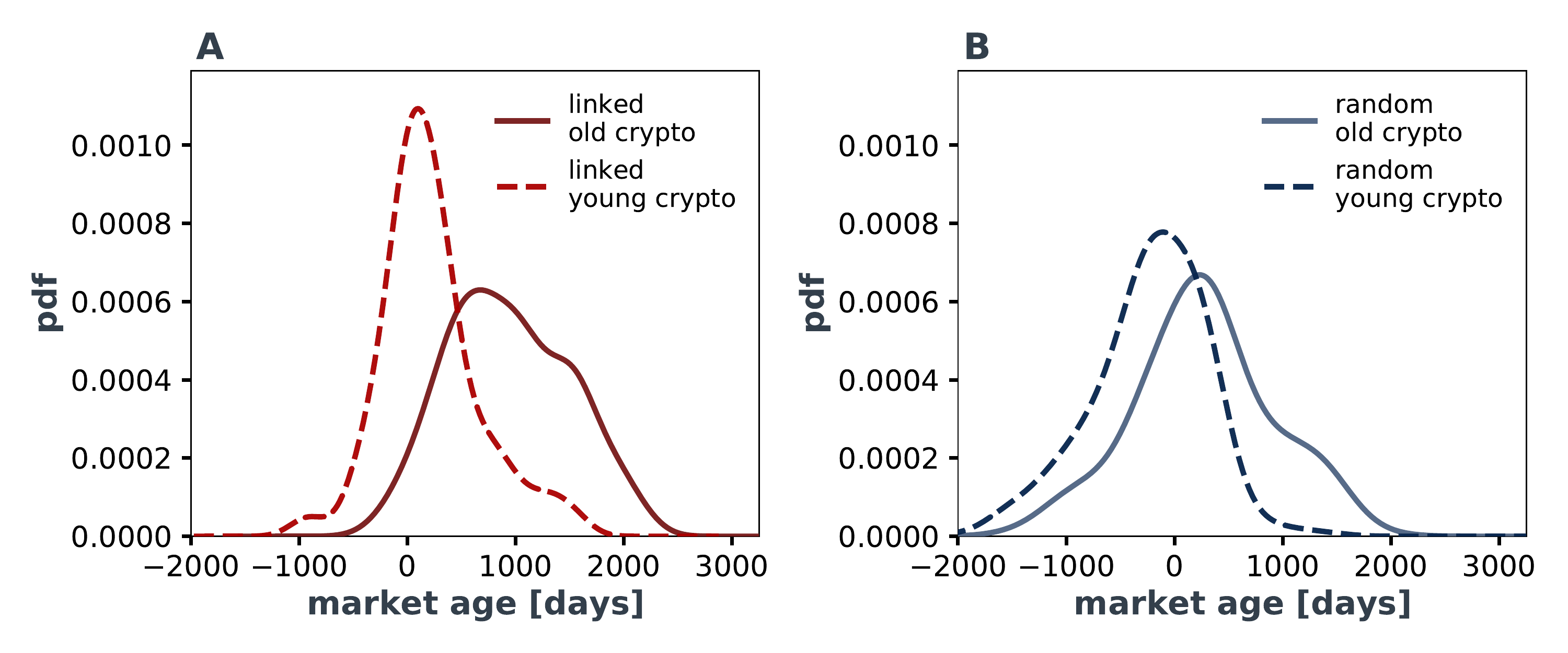}
	\label{SIfig15}
\end{figure}
\clearpage

%%% Fig. S16
\noindent {\bf Fig. S16.} \textbf{Connected cryptocurrencies are older and their age difference is greater than for random pairs.} The age (A) and age-difference (B) of cryptocurrencies whose correlation is increasing (continuous lines) and decreasing (dashed lines). Dot-dashed and dotted lines represent a set of random pairs whose correlation is increasing and decreasing, respectively.
\begin{figure}[h]
	\centering
	\includegraphics[width=0.99\linewidth]{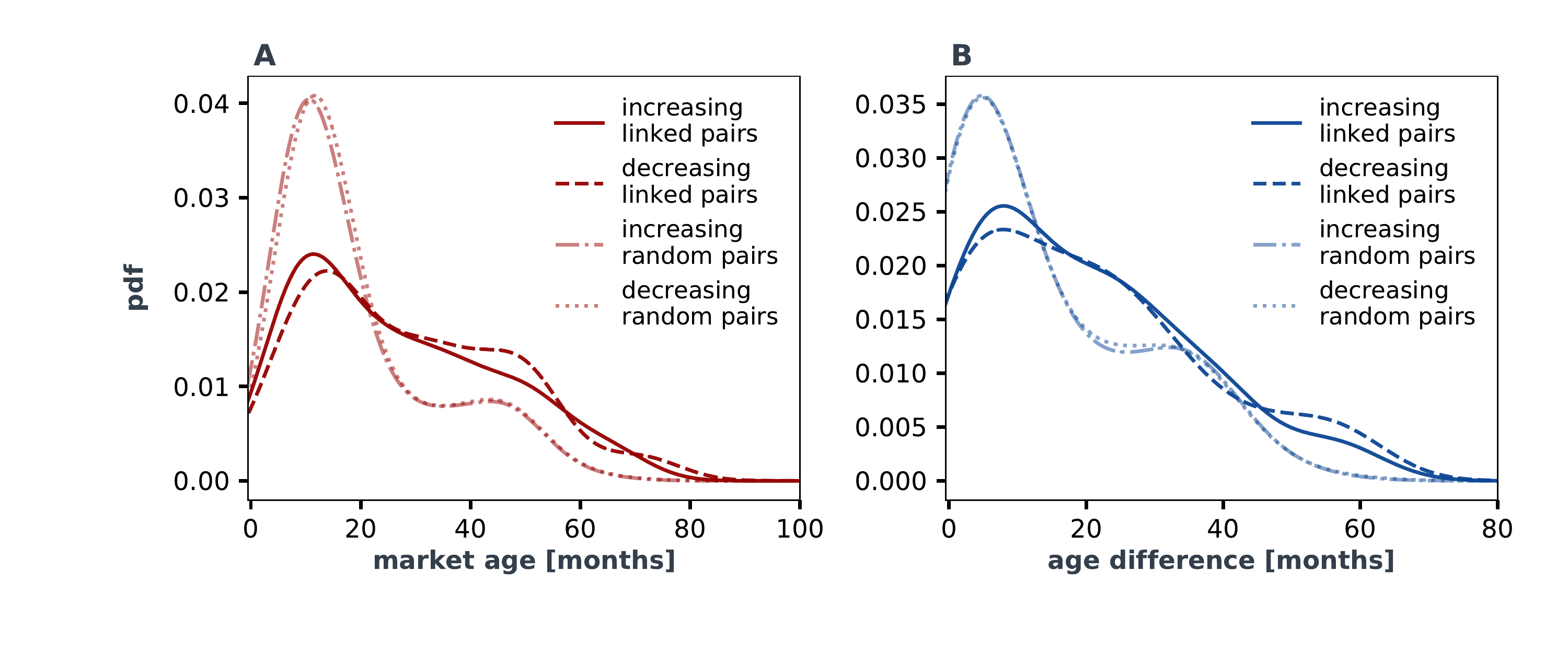}
	\label{SIfig16}
\end{figure}
\clearpage

%%% Fig. S17
\noindent {\bf Fig. S17.} \textbf{Comparing market capitalization and transaction volume of oldest and youngest currencies}. Market capitalization vs transaction volume for the oldest (red dots) and the youngest (blue dots) currency in each pair. Bars represent distributions.
\begin{figure}[h]
	\centering
	\includegraphics[width=0.65\linewidth]{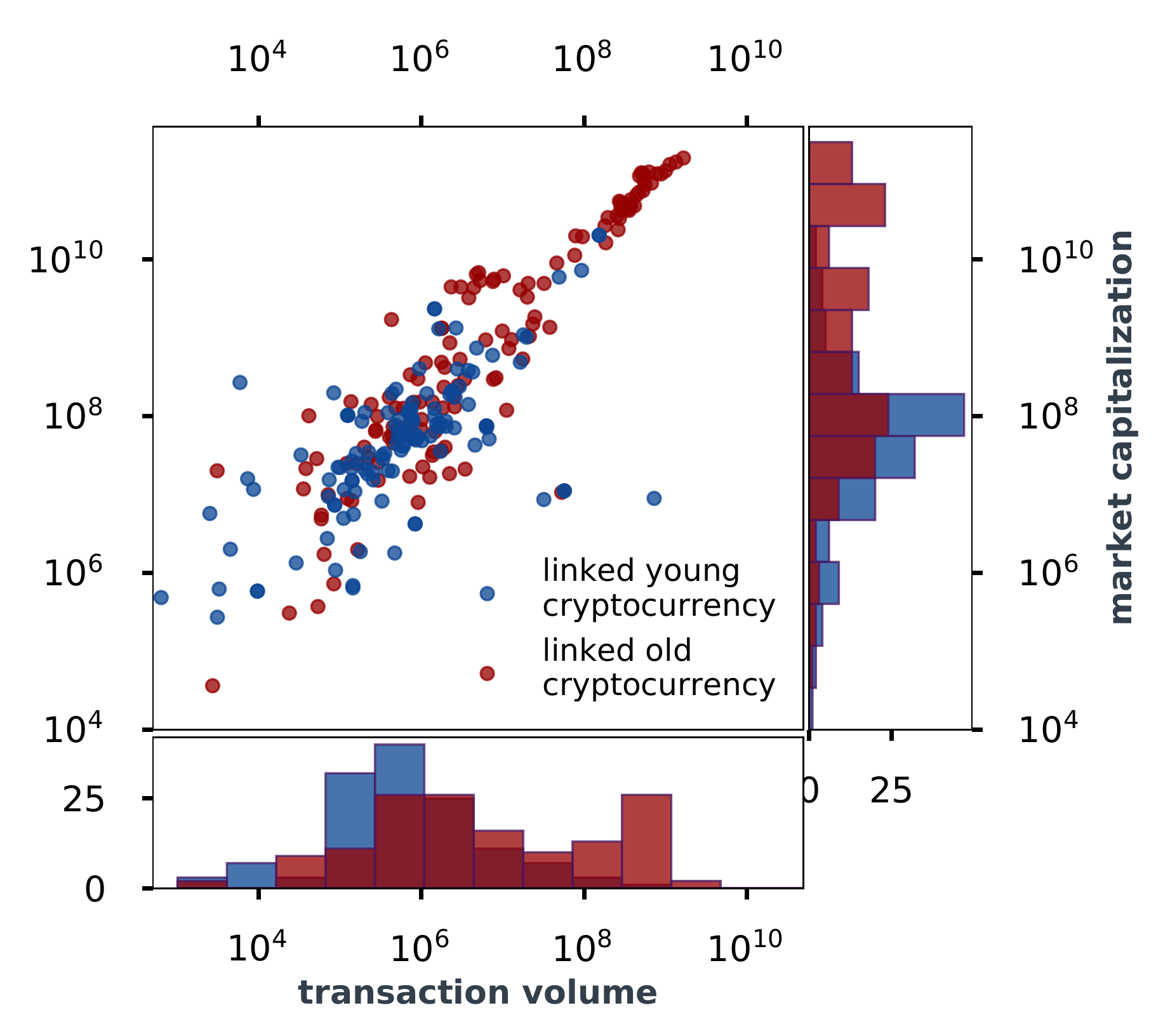}
	\label{SIfig17}
\end{figure}
\clearpage

%%% Fig. S18
\noindent {\bf Fig. S18.} \textbf{Linked cryptocurrencies differ in terms of market capitalization and volume}. The distribution of the market cap (A) and volume (B) difference between two cryptocurrencies in the same pair. Results are shown for cryptocurrency connected on GitHub before (dashed line) and after (filled line) the connection time, as well as for random pairs before (dotted-dashed line) and after (dotted line) the connection time.
\begin{figure}[t]
	\centering
	\includegraphics[width=0.99\linewidth]{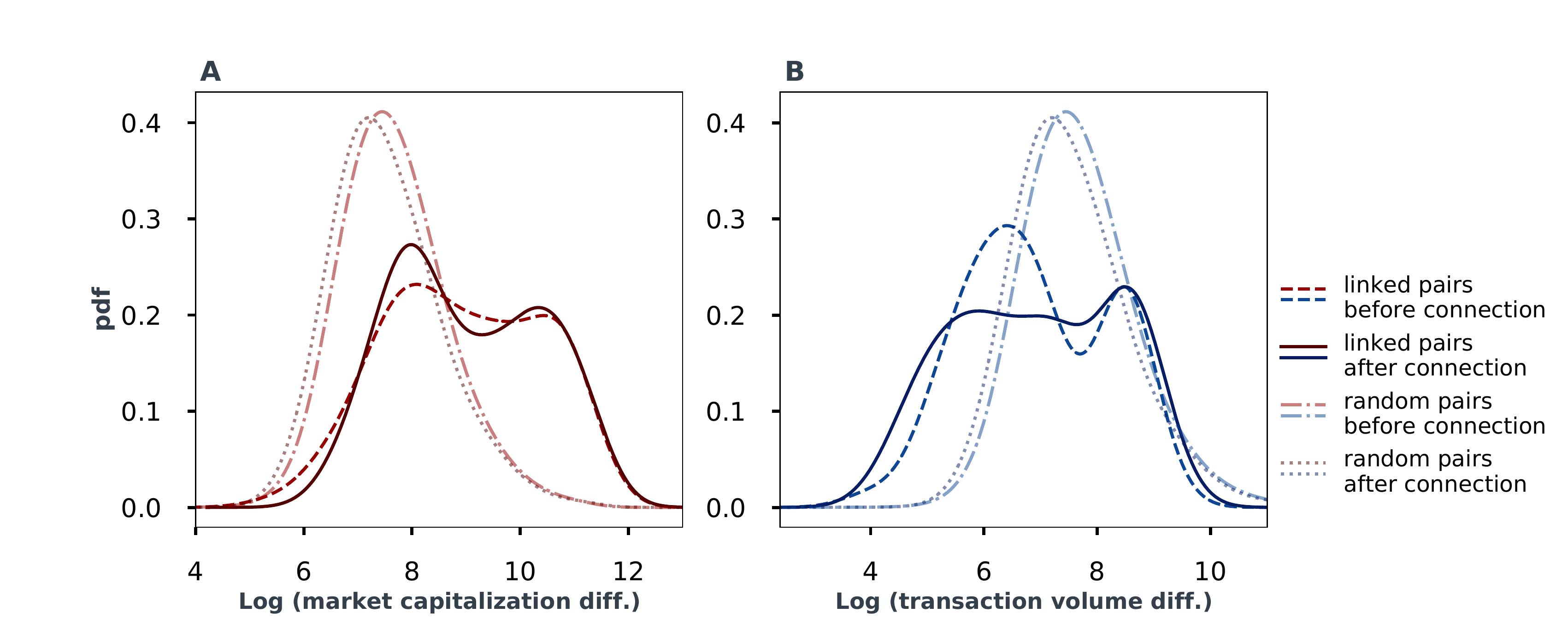}
	\label{SIfig18}
\end{figure}
\clearpage

%%% Fig. S19
\noindent {\bf Fig. S19.} \textbf{Market cap and volume: No difference between pairs with increasing vs decreasing market synchronization.} The distribution of the average market cap (A) and average volume (B) difference between two cryptocurrencies in the same pair. Results are shown for cryptocurrency connected on GitHub with increasing (dashed line) and decreasing (filled line) synchronization, as well as for random pairs before (dotted-dashed line) and after (dotted line) the connection time. Average market capitalization and average transaction volume are computed over the whole history of each currency.
\begin{figure}[t]
	\centering
	\includegraphics[width=0.99\linewidth]{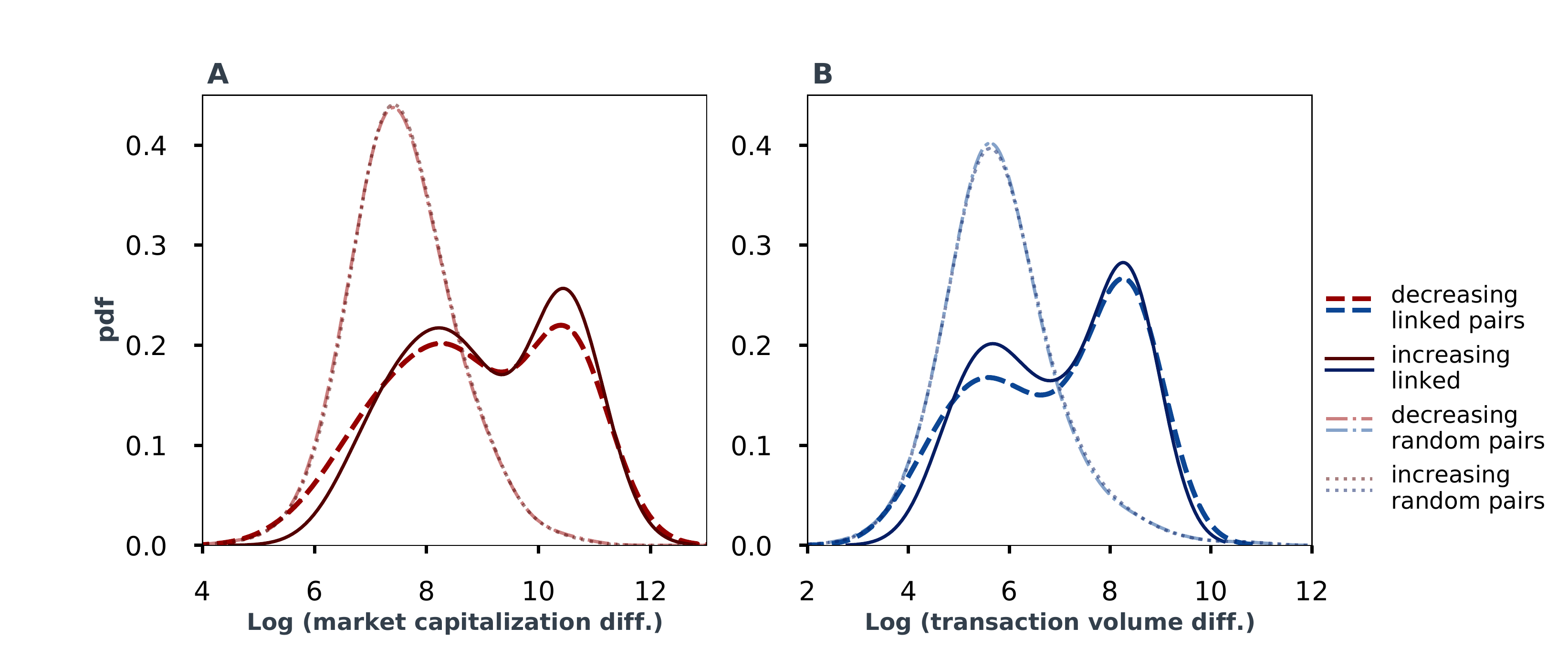}
	\label{SIfig19}
\end{figure}
\clearpage

%%% Fig. S20
\noindent {\bf Fig. S20.} \textbf{Market cap and volume: No difference between pairs with increasing vs decreasing market synchronization.} The distribution of the market cap (A) and volume (B) difference between two cryptocurrencies in the same pair. The difference is measured both before the connection for pairs whose correlation increases (dashed line) and decreases (filled line), and after the connection time for pairs whose correlation increases (dashed-dotted line) and decreases (dotted line). 
\begin{figure}[t]
	\centering
	\includegraphics[width=0.99\linewidth]{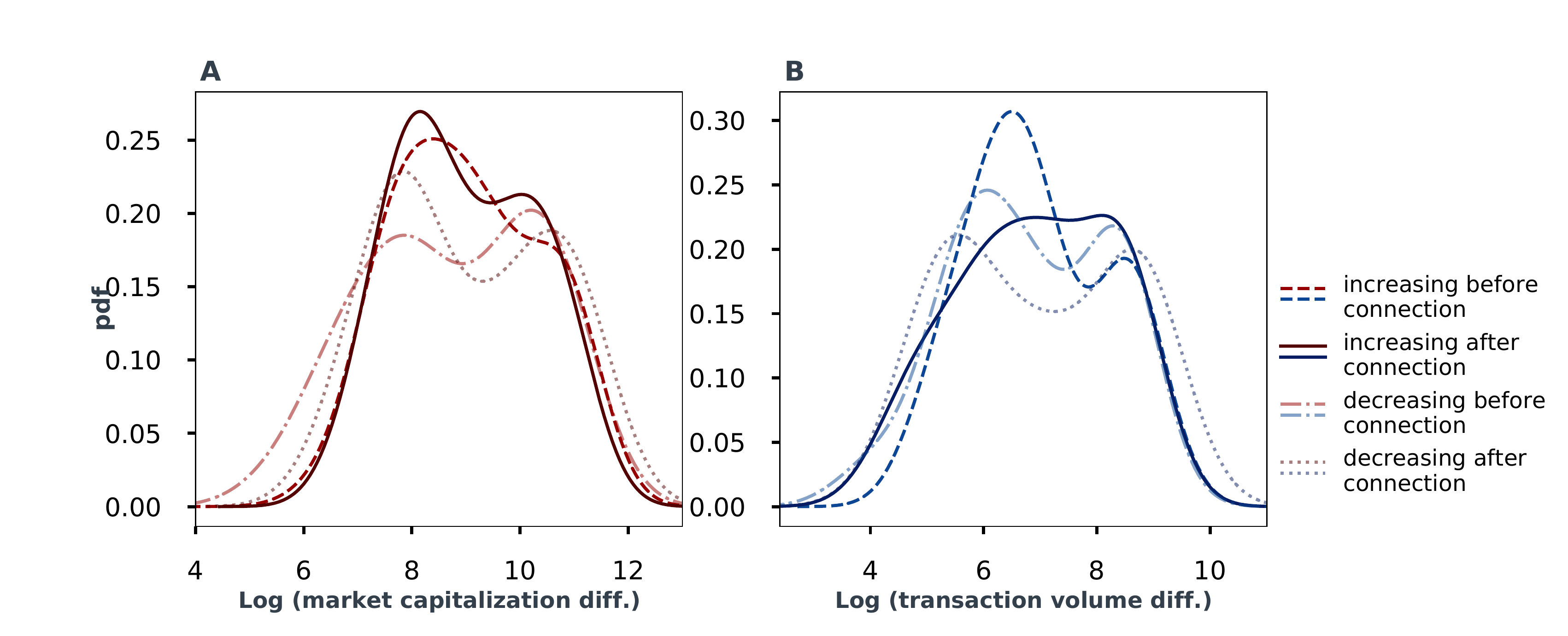}
	\label{SIfig20}
\end{figure}
\clearpage

%%% Fig. S21
\noindent {\bf Fig. S21.} \textbf{Which pairs of linked cryptocurrencies synchronize?} The standardised correlation after vs before the connection time for two different classes of linked pairs (blue circles and red pentagons). Results are shown for the (A) young-first vs young-second class, (B) high-diff vs low-diff class, (C) keep vs dismiss class, (D) top vs minor class, (E) push vs pull class, and (F) late vs early class (see text for definitions). The straight line corresponds to the case of no change in correlation. In panels (C) and (D) the difference between the two classes is significant, e.g. under Welch's test for average, Mood's and Kruskal-Wallis tests for the median, Mann-Whitney U, and Kolmogorov-Smirnov tests.
\begin{figure}[t]
	\centering
	\includegraphics[width=0.8\linewidth]{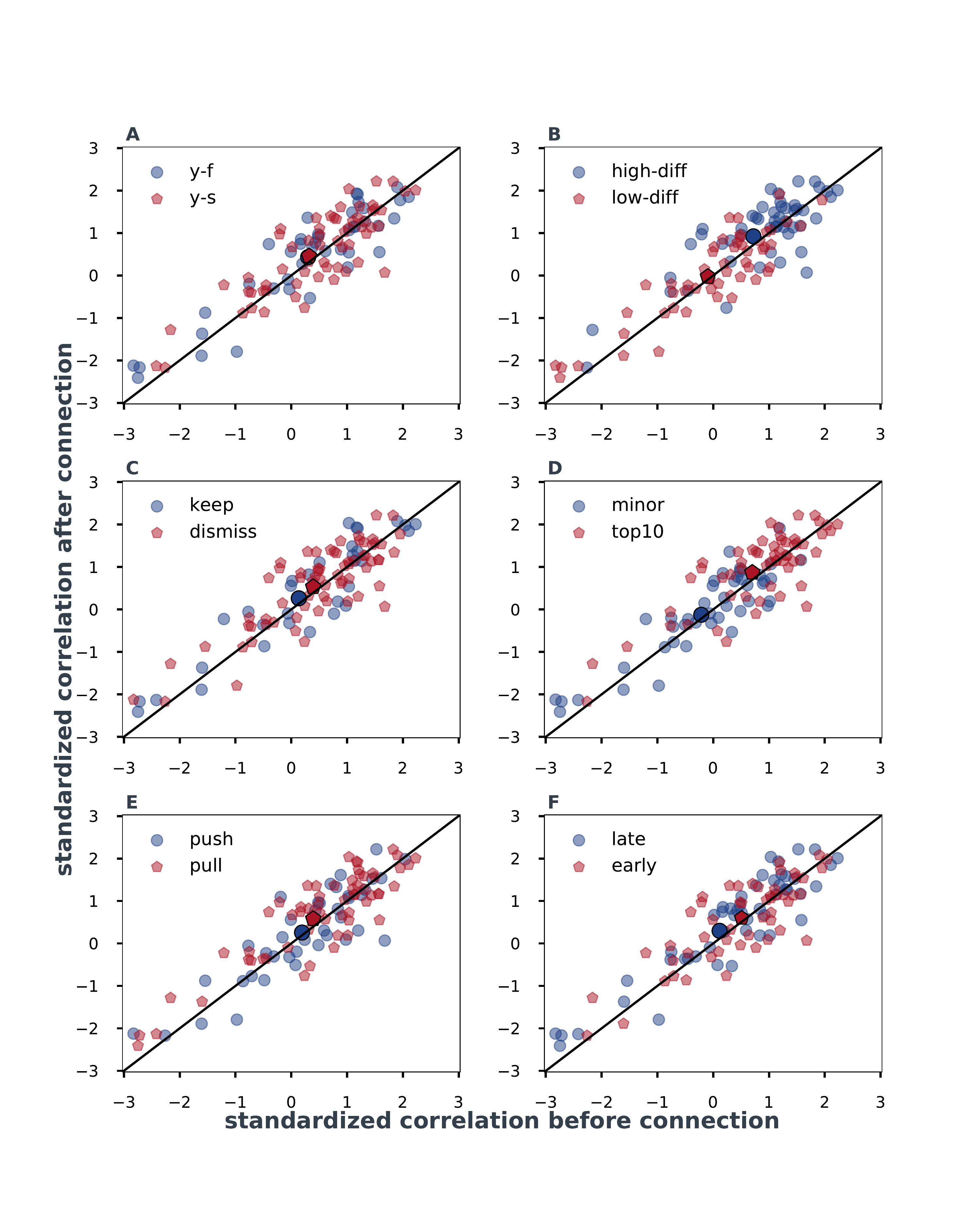}
	\label{SIfig21}
\end{figure}
\clearpage

\end{document}